\documentclass[pdflatex,sn-nature]{sn-jnl}

\usepackage{amssymb}
\usepackage{graphicx}%
\usepackage{multirow}%
\usepackage{amsmath,amssymb,amsfonts}%
\usepackage{amsthm}%
\usepackage{mathrsfs}%
\usepackage[title]{appendix}%
\usepackage{xcolor}%
\usepackage{textcomp}%
\usepackage{manyfoot}%
\usepackage{booktabs}%
\usepackage{algorithm}%
\usepackage{algorithmicx}%
\usepackage{algpseudocode}%
\usepackage{listings}%
\usepackage{float} 

\theoremstyle{thmstyleone}%
\theoremstyle{thmstyletwo}%

\theoremstyle{thmstylethree}%

\begin{document}

\title[Article Title]{A Linear Generative Framework for Structure-Function Coupling in the Human Brain}


\author[1]{\fnm{Sam Frank} \sur{Kelemen}}

\author[2]{\fnm{Joaquín} \sur{Goñi}}

\author[3]{\fnm{S\'ergio} \sur{Pequito}}

\author[1]{\fnm{Arian} \sur{Ashourvan}}\email{ashourvan@ku.edu}

\affil[1]{\orgdiv{Department of Psychology}, \orgname{University of Kansas}, \orgaddress{\street{Jayhawk Boulevard}, \city{Lawrence}, \postcode{66045}, \state{Kansas}, \country{USA}}}

\affil[2]{\orgdiv{Edwardson School of Industrial and Weldon School of Biomedical Engineering}, \orgname{Purdue University}, \orgaddress{\street{206 S Martin Jischke Dr,}, \city{West Lafayette}, \postcode{47907}, \state{Indiana}, \country{USA}}}

\affil[3]{\orgdiv{Institute for Systems and Robotics}, \orgname{Instituto Superior Técnico, Universidade de Lisboa}, \orgaddress{ \city{Lisbon}, \postcode{66045},  \country{Portugal}}}



\abstract{ Brain function emerges from coordinated activity across anatomically connected regions, where structural connectivity (SC) -- the network of white matter pathways -- provides the physical substrate for functional connectivity (FC) -- the correlated neural activity between brain areas. While these structural and functional networks exhibit substantial overlap, their relationship involves complex, indirect mechanisms, including the dynamic interplay of direct and indirect pathways, recurrent network interactions, and neuromodulatory influences. To systematically untangle how structural architecture shapes functional patterns, this work aims to establish a set of \emph{rules} that decode how direct and indirect structural connections and motifs give rise to FC between brain regions. Specifically, using a generative linear model, we derive explicit rules that predict an individual's resting-state fMRI FC from diffusion-weighted imaging (DWI) -- derived SC, validated against topological null models. Examining the rules reveals distinct classes of brain regions, with \emph{integrator} hubs acting as structural linchpins promoting synchronization and \emph{mediator} hubs serving as structural fulcrums orchestrating competing dynamics. Through virtual lesion experiments, we demonstrate how different cortical and subcortical systems distinctively contribute to global functional organization. Together, this framework disentangles the mechanisms by which structural architecture drives functional dynamics, enabling the prediction of how pathological or surgical disruptions to brain connectivity cascade through functional networks, potentially leading to cognitive and behavioral impairments.}




\maketitle

\section{Introduction}\label{sec1}

Understanding how brain function emerges from anatomical structure represents one of the central challenges in neuroscience.  While the neuron doctrine established the foundational principle that neural activity propagates through synaptic connections \cite{shepherd2015foundations}, translating this insight to explain large-scale brain dynamics across millions of neurons and complex network architectures remains a formidable challenge. However, this foundational principle scales to extraordinary complexity when considering the stochastic dynamics of neuronal noise, the intricate integration across thousands of dendritic inputs \cite{london2005dendritic}, and the hierarchical organization of connectivity patterns that emerge across millions of interconnected neurons spanning multiple spatial and temporal scales.

Given this multi-scale complexity, no unified model currently bridges the dynamics of individual neurons to whole-brain network interactions. Consequently, neuroscience has adopted a levels-based approach, with macroscopic models treating brain regions as abstract nodes governed by phenomenological or biophysically-informed dynamics~\cite{deco2008dynamic}. Within this framework, efforts to link structural connectivity (SC) to functional connectivity (FC) have crystallized into three distinct modeling paradigms \cite{suarez2020linking}:  (i) \emph{statistical approaches} utilize pattern recognition and correlation maximization to identify SC-FC correspondences \cite{smith2013functional,mivsic2016network,messe2014relating}, but remain primarily descriptive in revealing areas of convergence and divergence between modalities; (ii) \emph{mechanistic models}, including dynamic systems \cite{honey2007network,tanner2022redefining} and biophysical frameworks \cite{breakspear2017dynamic,sanz2015mathematical,deco2009key}, simulate the propagation of activity through structural networks to explain the dynamics of the emergent brain, though their interpretations depend heavily on the underlying assumptions about the dynamics of the neural population and the linearity between regions; and (iii) \emph{communication models} bridge these approaches by leveraging network science and graph theory to understand how topology shapes information flow \cite{mišic2014communication,mivsic2015cooperative,graham2011packet,crofts2009weighted,becker2018spectral}, operationalizing FC through concepts like shortest-path routing \cite{goni2014resting}, path ensembles \cite{crofts2009weighted,avena2018communication}, and transmission-diffusion or coactivation mechanisms \cite{abdelnour2014functional,mivsic2015cooperative}.

In this work, we present a novel approach that combines elements of statistical and communication models. Inspired by the seminal work of Barabasi et al. \cite{barabasi2020genetic,kovacs2020uncovering}, who proposed a generative framework to predict SC in \emph{C. elegans} worms based on their genetic profile, we leverage a similar framework to predict FC in the human brain. Their model utilizes a set of empirically derived \emph{rules} that act linearly on the genetic fingerprint to predict pairwise structural connections between regions. We adapt this framework to predict resting-state fMRI FC between brain regions based on their structural fingerprint measured by diffusion-weighted MRI (DWI). By doing so, we aim to demonstrate how both direct and higher-order connection patterns and motifs between neurons contribute to the observed patterns of functional connectivity. Subsequently, we will utilize the derived rules to explore the influence of various cortical and subcortical regions on large-scale correlation and anti-correlation patterns across the cortex.

\section{Results}\label{sec2}

\begin{figure*}
\centering
\includegraphics[width=1\linewidth ]{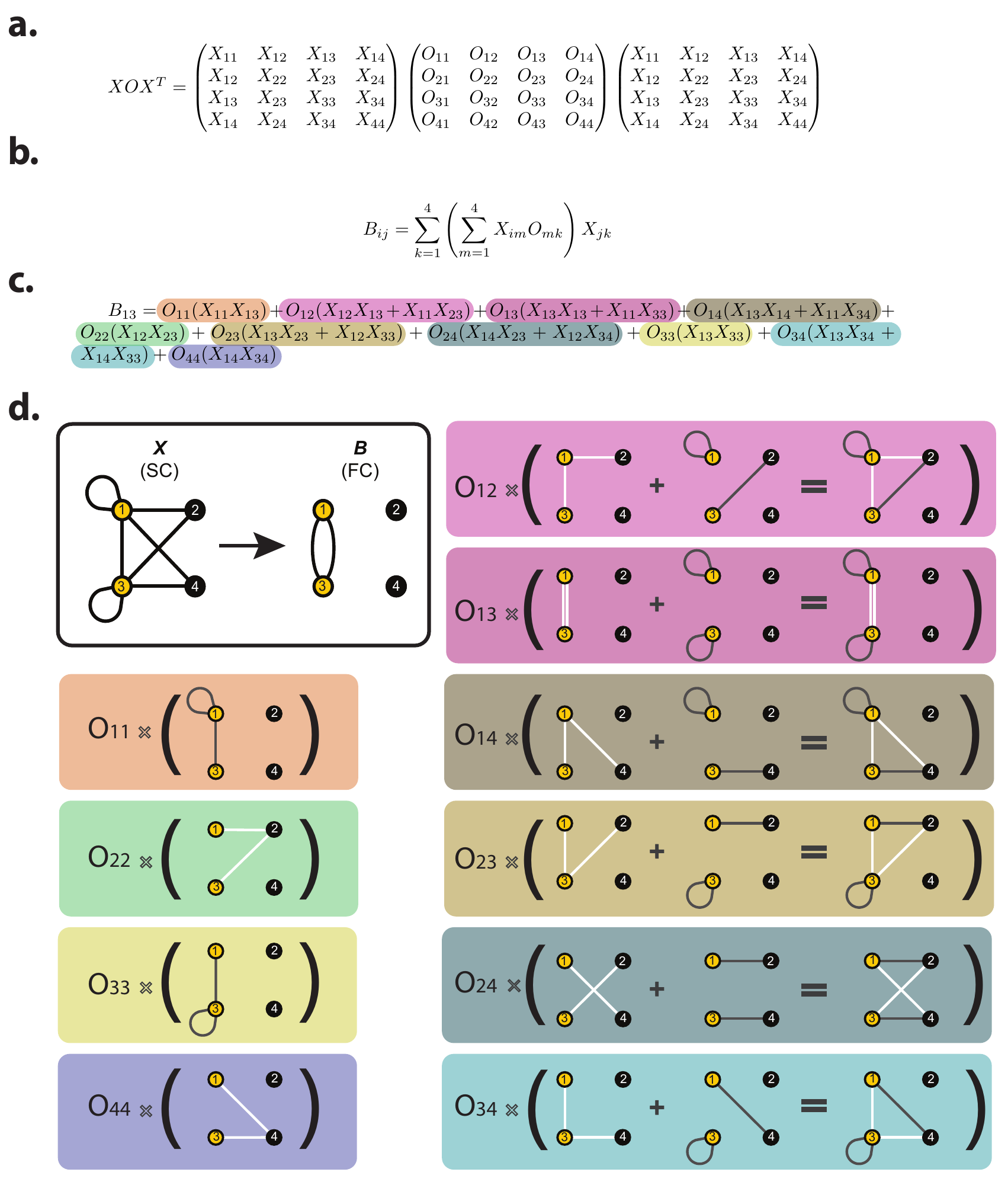}
\caption{\textbf{Network motifs and connectivity patterns captured by structural generative rules.} 
\textbf{(a)} The equation for generating the $4\times 4$ symmetric FC matrix ($B$) from the SC matrix ($X$) using the generative rule matrix ($O$). Since $O$ is symmetric (see the Materials and Methods for details), the element-wise equation for $B$ can be represented as depicted in panel \textbf{b}. \textbf{(c)} Expanded equation illustrating the generation of FC between nodes 1 and 3 using the corresponding elements of $X$ and $O$. The color-coded sections highlight various connection patterns and motifs, as depicted in panel \textbf{d}, contributing to the FC between nodes 1 and 3.}
\label{Figure_1}
\end{figure*}



Let FC be represented by the symmetric matrix $B \in \mathbb{R}^{N \times N}$ and SC by the symmetric matrix $X \in \mathbb{R}^{N \times N}$, where $N$ is the number of brain regions. We propose a linear generative model given by the equation $B = XOX^T$, where $O \in \mathbb{R}^{N \times N}$ is the symmetric \emph{rule} matrix that captures how SC patterns with other regions influence the FC weight between two regions. In Fig. \ref{Figure_1}, we illustrate how the elements of the rule matrix encode various direct and monosynaptic structural connection patterns and how their linear combinations contribute to the FC weight between two regions. 

\subsection{Predicting FC from individuals' SC }

\subsubsection*{Subject-level rule estimation and prediction accuracy}


In Fig. \ref{Figure_2}, we present the estimated rule matrix and the predicted FC values for a sample subject using LASSO regression -- see Materials and Methods for details. The model demonstrates high accuracy in predicting FC values, as evidenced by the close linear fit between the original and predicted values shown in Fig. \ref{Figure_2}e. Across all subjects, the model achieves an average slope of \( 0.94 \pm 0.04 \) and an average \( R^2 \) of \( 0.81 \pm 0.05 \).

\subsubsection*{Group-level rule estimation reveals universal principles}

To uncover universal rules for predicting FC from subjects' SC patterns and the extent of intersubject variability, we fit a single rule matrix \( O \) to all subjects' SC and FC matrices at the group level. We leveraged the Kronecker product and inverted the resulting \mbox{group-level} SC matrix -- see the Materials and Methods for details. 

\begin{figure*}
\centering
\includegraphics[width=1\linewidth ]{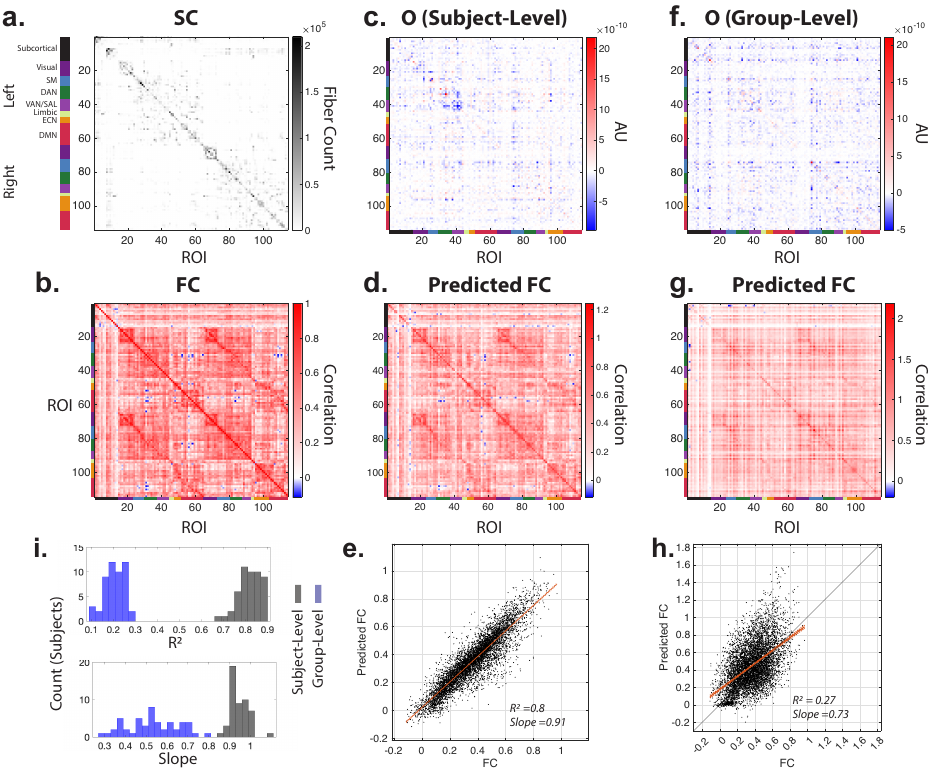}
\caption{\textbf{Predicted FC from SC.} 
\textbf{(a)} Sample subject's SC matrix and \textbf{(b)} corresponding FC matrix.  \textbf{(c)} The estimated subject-level rule matrix, highlighting significant values compared to a randomized null model with preserved degree distribution ($p<0.05$, FDR corrected for multiple comparisons), and \textbf{(d)} the predicted FC matrix using the rules from panel \textbf{c}.  \textbf{(e)} Scatter plot comparing the sample subject's actual FC values with the predicted FC values, and the red line indicates the linear fit. \textbf{(f)} The estimated group-level rule matrix, highlighting significant values compared to a randomized null model, and \textbf{(g)} the predicted FC matrix for the same sample subject using the group-level rules from panel \textbf{f}. Here, the predicted FC values equal to or higher than one are represented with the same color to aid visualization. \textbf{(h)} Scatter plot comparing the sample subject's actual FC values with the predicted FC values using the group-level rule matrix from panel \textbf{f}. The red line shows the linear fit, and the dashed line the $95\%$ Confidence interval. \textbf{(i)} Distributions of the $R^2$ values (top) and slopes of the linear fit between the actual and predicted FC values using the subject- (black) and \mbox{group-level} (blue) models.}
\label{Figure_2}
\end{figure*}

In Fig. \ref{Figure_2}g, we show the predicted FC for a sample subject using the group-level rule matrix presented in Fig. \ref{Figure_2}f. The linear fit between the empirical and predicted FC values in Fig. \ref{Figure_2}h indicates that group-level rules can indeed generate FC, with an average slope of \( 0.53 \pm 0.12 \) and \( R^2 \) of \(0.2 \pm 0.04 \) across all subjects. However, compared to the \mbox{subject-level} rules, the group approach results in a significant ($t-$test,  $p=2\times 10^{-48}$) reduction in the goodness-of-fit of the linear regression fit between empirical and predicted FC.

\subsubsection*{Rule matrix sparsity and robustness analysis}

\begin{figure*}
\centering
\includegraphics[width=1\linewidth ]{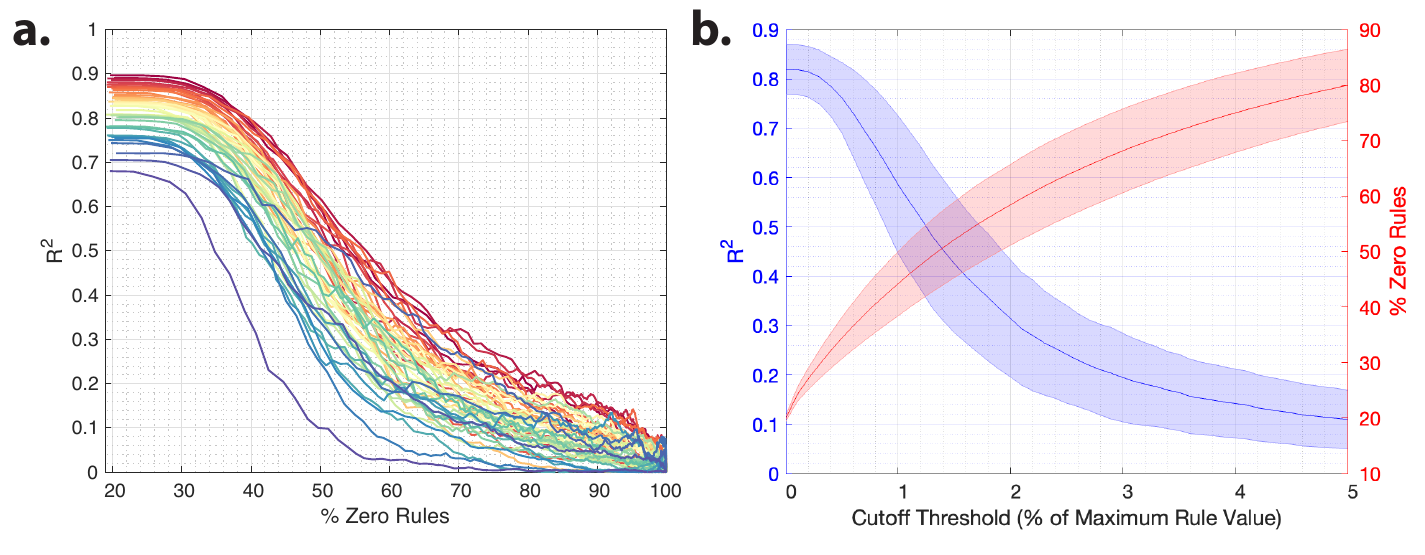}
\caption{\textbf{Impact of thresholding the rule matrix on prediction accuracy} \textbf{(a)} The model's accuracy is measured by the goodness-of-fit of a linear regression fit, indicated by $R^2$, comparing the actual versus predicted FC values. Each color-coded line represents the $R^2$ value associated with each subject as we increase the threshold, replacing the elements of the rule matrix $O$ below the cutoff with zero. The x-axis represents the total percentage of zero elements in the rule matrix as we increase the threshold. \textbf{(b)}The shaded blue line indicates the averaged $R^2$ values across subjects, while the thickness of the shaded area reflects the standard deviation. The x-axis displays the cutoff threshold values as percentages, relative to the maximum rule element value. The red line represents the average proportion of zero elements relative to the total number of elements in percentage at each cutoff threshold.}
\label{figure_3}
\end{figure*}

The rule matrix exhibits a heavy-tailed weight distribution, with a high density of zero values -- see Supplementary Information (SI) Fig. \ref{SI_Figure_1}. To assess the sensitivity of the model predictions to these smaller-weighted rules, we incrementally replaced the lowest-valued rules with zero and evaluated the resulting prediction accuracy (Fig. \ref{figure_3}). The prediction accuracy remained relatively stable until the group-average proportion of zero elements reached approximately 30\% of the total rules (Fig. \ref{figure_3}b), after which it sharply declined.

\subsubsection*{Subject-specific parameters do not generalize across individuals}

Human brain networks have universal properties (small-world, modular, heavy-tailed degree distributions) despite individual structural differences. To investigate how a subject's unique structural network influences the prediction of functional connections, we compared predictions made using their own SC  with those made using another subject's SC, both using the same model parameters.

Our results indicate that the subject-specific model parameters are highly individualized and do not generalize well to other subjects' SC (SI Fig.\ref{SI_Figure_3_new}). We calculated the predictive accuracy of the model by performing a linear regression of the predicted versus observed FC values. The significant drop ($t-$test, $p=8.1 \times 10^{-61}$) in model accuracy (i.e., $R^{2}$) when predicting FC using another subject's SC, compared to using the subject’s own, underscores this finding (SI Fig.\ref{SI_Figure_3_new}a). 

For each subject, we also evaluated the predictive accuracy of the model when we swapped the subject's SC with that of edge-randomized SC null models, preserving the degree and strength distributions of each brain region. Similarly, the predictions are highly inaccurate when we replace the SC with SC nulls  ($t-$test, $p=8.1 \times 10^-61$). 

\subsubsection*{Topological features preserve predictive performance}

We repeated the previous analyses, but with a change in how the rule matrices ($O$) were estimated. Instead of using each subject's own SC and FC matrices, we estimated $O$ using another subject's SC matrices. Our results indicate that fitting the model parameters leads to highly accurate predictions, even when using data from another subject's SC or null SCs (SI Fig. \ref{SI_Figure_3_new}c-d). Statistical testing confirmed that the results are indistinguishable at the group level. Supplementary analysis at the individual level revealed an inconsistent pattern of performance that was either comparable to, lower than, or higher than the null expectations across subjects, as shown in SI Fig. \ref{SI_Figure_3}.  


\subsubsection*{Statistical significance of rule elements}

We assessed the model's accuracy after retaining only the rules whose values exceeded those estimated from the edge-randomized SC null model. To identify significant rule elements at the subject-level, we used nonparametric permutation testing ($p<0.05$, FDR-corrected for multiple comparisons) using $n=100$ randomized SC null models. On average, more than 4$\%$ of rule elements did not pass the significance level (SI Fig.~\ref{SI_Figure_2}a). Importantly, by removing the non-significant elements results in a small, yet significant ($t-$test, $p=1.01 \times 10^{-14}$), reduction in the model prediction accuracy (SI Fig. \ref{SI_Figure_2}b). Together, these results indicate that preserved topological features, such as the degree sequence, play a crucial role in shaping functional connectivity predictions. 

\subsubsection*{Direct versus indirect structural features in FC prediction}

We compared the model’s predicted FC using SC with predictions based on search information \citep{Goni2014}, a metric that captures the navigability of shortest paths across the network -- see the Materials and Methods for details. This comparison allows us to assess whether functional interactions are better explained by direct anatomical links or by the ease of communication through the broader network topology, which is an important distinction for identifying the mechanisms underlying functional coupling. 

Our results show that models driven by each subject’s own SC achieve markedly higher predictive accuracy, as reflected in significantly larger $R^2$ values (SI Fig. \ref{SI_Figure_4_SI}). This implies that the direct, monosynaptic connections represented in the SC already account for the bulk of the FC variance. While search information can capture part of the residual variance, replacing SC with the search information matrix as input in our generative model diminishes its ability to represent the effect of the immediate structural links accurately.

\subsection{Organizing the brain by structure-function coupling roles}

\begin{figure*}[hbtp]
\centering
\includegraphics[width=1\linewidth ]{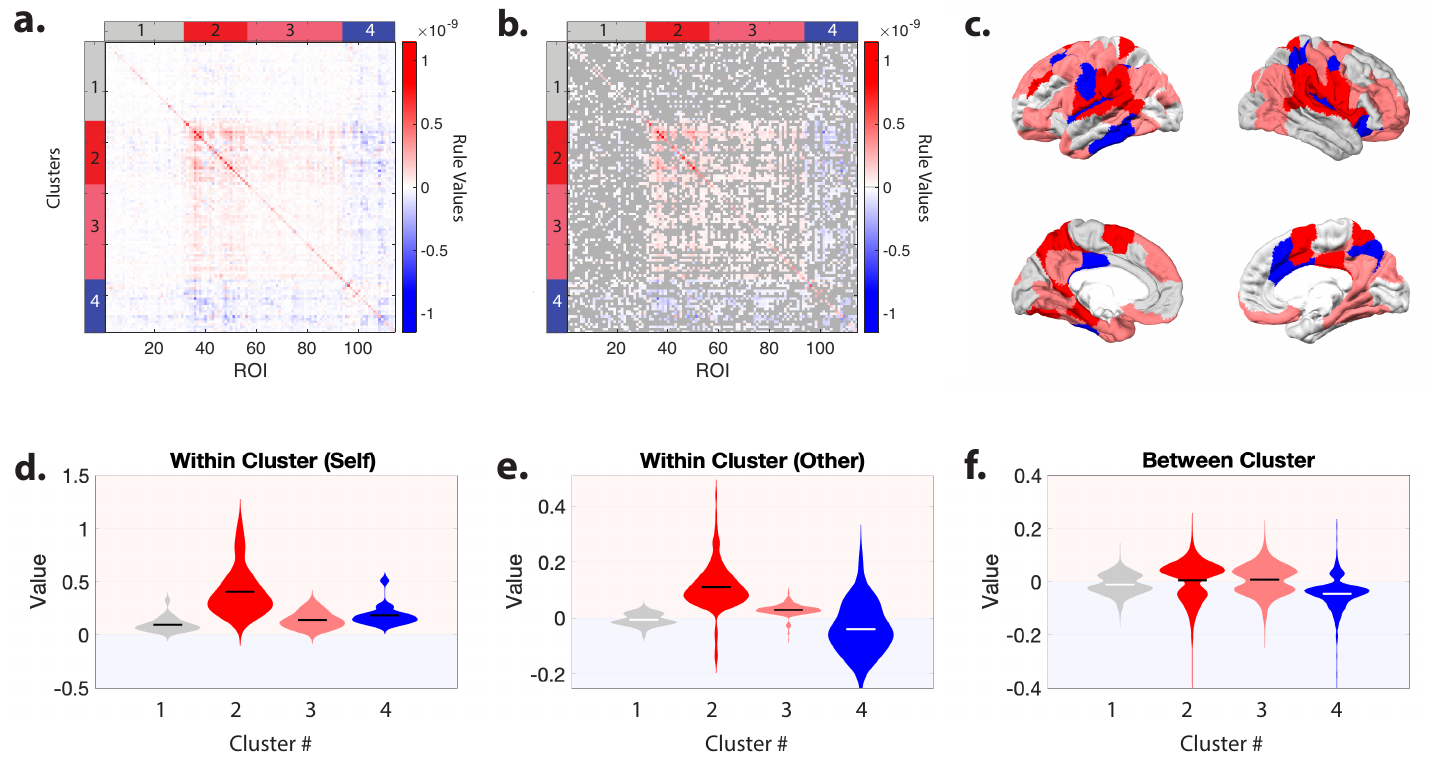}
\caption{\textbf{Structural linchpins and fulcrums of functional dynamics.} \textbf{(a)} Subject-level rule matrix $O$ averaged across all subjects, with brain regions sorted by their cluster assignments identified using the WSBM method ($k=4$). \textbf{(b)} Same matrix in panel \textbf{a}, except the rule elements with means that show no significant ($t-$test, $p<0.05$, FDR-corrected for multiple comparisons across all rules) difference from zero across all subjects are color-coded in gray  \textbf{(c)} Brain overlay of the identified clusters. Distributions of the significant within-cluster self-connections (diagonal elements) \textbf{(d)}, within-cluster other connections (off-diagonal elements) \textbf{(e)}, and between-cluster connections \textbf{(f)} from the matrix in panel \textbf{b}, color-coded according to their community assignments. Black (white) bars show the means of distributions with positive (negative) signs.}
\label{Figure_4}
\end{figure*}

\begin{figure*}
\centering
\includegraphics[width=1\linewidth ]{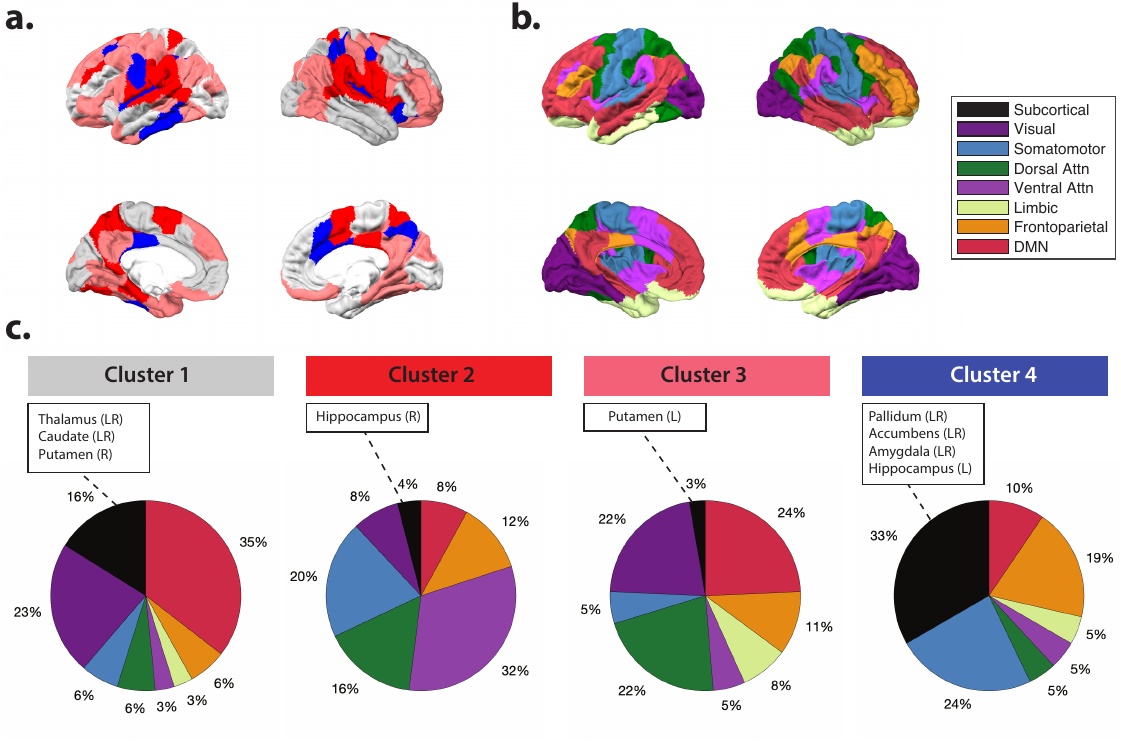}
\caption{\textbf{Rule matrix clusters overlap with resting-state networks.} 
\textbf{(a)} Brain overlay of clusters identified from the mean rule matrix $O$ in Fig. \ref{Figure_4}, juxtaposed with the seven resting-state brain networks. 
\textbf{(b)} Resting-state networks as identified by Yeo and colleagues \cite{thomas2011organization}. 
\textbf{(c)} Percentage contribution of resting-state ROIs to the identified clusters.}
\label{Figure_5}
\end{figure*}

\subsubsection*{Disentangling direct and structural motif contributions to FC}

The rule matrix $O$ quantifies how direct SC and first-order indirect SC pathways collectively shape FC. The prevalence of high diagonal values across all regions highlights the importance of direct SC and first-order indirect pathways in shaping FC. These findings remain robust even after removing non-significant rules at the subject level, with only an approximately $2\%$ difference in the identification of mean significant rules across all subjects (SI Fig. \ref{SI_Figure_2}). 
However, the presence of off-diagonal weights implies that structural motifs (see Fig. \ref{Figure_1}d) also contribute to FC. 

\subsubsection*{Rule matrix organization reveals structural linchpins and fulcrums shaping FC} 

To elucidate the structural roles of brain regions in driving these dynamics, we applied weighted stochastic block modeling (WSBM) -- see the Materials and Methods for details --  to partition the $O$ averaged across all subjects into four clusters based on similarities in their within- and between-cluster connection profiles (Fig. \ref{Figure_4}). 

Our approach revealed distinct organizational principles: Cluster 1 exhibited weak off-diagonal interactions. Cluster 2 and, to a lesser extent, Cluster 3 displayed positive intra- and inter-cluster weights, indicative that contributions from indirect SC motifs enhance FC. 
In contrast, Cluster 4 was uniquely characterized by globally negative off-diagonal weights, suggesting structural motifs involving this cluster's brain regions suppress FC. 

Additionally, we mapped each cluster’s ROIs to seven canonical resting-state networks and subcortical regions (Fig. \ref{Figure_5}). This analysis revealed heterogeneous contributions across clusters. For example, nearly half of the regions of interest in Cluster~2 were from attention networks, while Cluster 4 primarily comprised subcortical, somatomotor, and frontoparietal ROIs.

Together, these results show that WSBM-based decomposition of the rule matrix $O$ delineates distinct structural classes, such as \textit{mediator} (Cluster 4) and \textit{integrator hubs} (Clusters 2-3), whose motif-mediated structural interactions shape large-scale functional dynamics. Integrator hubs serve as \emph{structural linchpins}, promoting synchronization among the regions they connect, while mediator hubs function as \emph{structural fulcrums}, facilitating competitive dynamics between their connected regions. Together, these findings provide a taxonomy of brain regions based on how structural specialization shapes emergent functional connectivity in the brain.


\subsection{Examining the impact of regional structural profiles on whole-brain FC}

Our generative framework provides a principled way to tease apart how individual regions or networks shape whole-brain FC. We accomplish this by retaining only those SC edges that originate from or terminate in a chosen target region or network. Leveraging the group-level rule matrix $O$, we demonstrate how the framework quantifies each region’s specific contribution to global FC. As a proof of concept, we simulate the effect of selectively preserving only the default mode network (DMN) structural connections (Fig. \ref{Figure_6}), demonstrating that our model yields viable and informative predictions. Fig. \ref{Figure_6} illustrates the group-average contribution of the default mode network (DMN) to FC. The DMN’s SC not only directly shapes intra-DMN FC and the coupling between DMN nodes and extra-DMN regions, but also indirectly modulates FC among pairs of non-DMN regions via their shared connections with DMN ROIs. This demonstration highlights how our framework can be used to systematically assess the impact of specific regions and networks on global FC. More importantly, it underscores the model’s utility for studying how localized structural disruptions, such as those resulting from aging, neurodevelopmental disorders, traumatic brain injury, or surgical resection, alter whole-brain functional organization.

\begin{figure*}
\centering
\includegraphics[width=1\linewidth ]{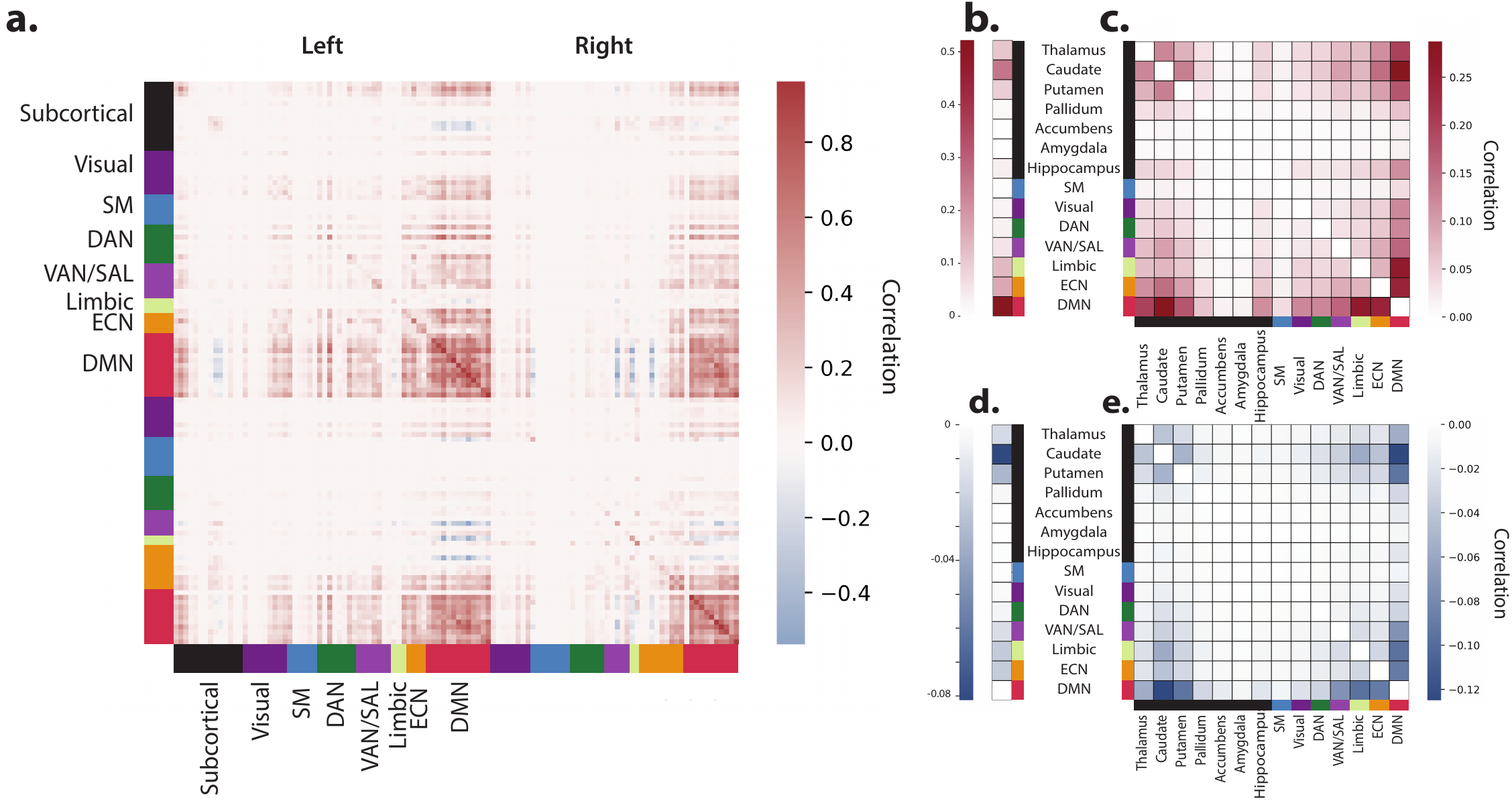}
\caption{\textbf{Default mode network's influence on global FC patterns.} 
 \textbf{(a)} Predicted FC matrices after selectively removing the influence of all regions except for the DMN regions of interest. 
\textbf{(b)} Average predicted positive FC values within each ROI for different systems, and \textbf{(c)} between different ROIs for different systems. 
\textbf{(d)} Average predicted negative FC values within each ROI for different systems, and \textbf{(e)} between different ROIs for different systems.}
\label{Figure_6}
\end{figure*}

\begin{figure*}
\centering
\includegraphics[width=1\linewidth ]{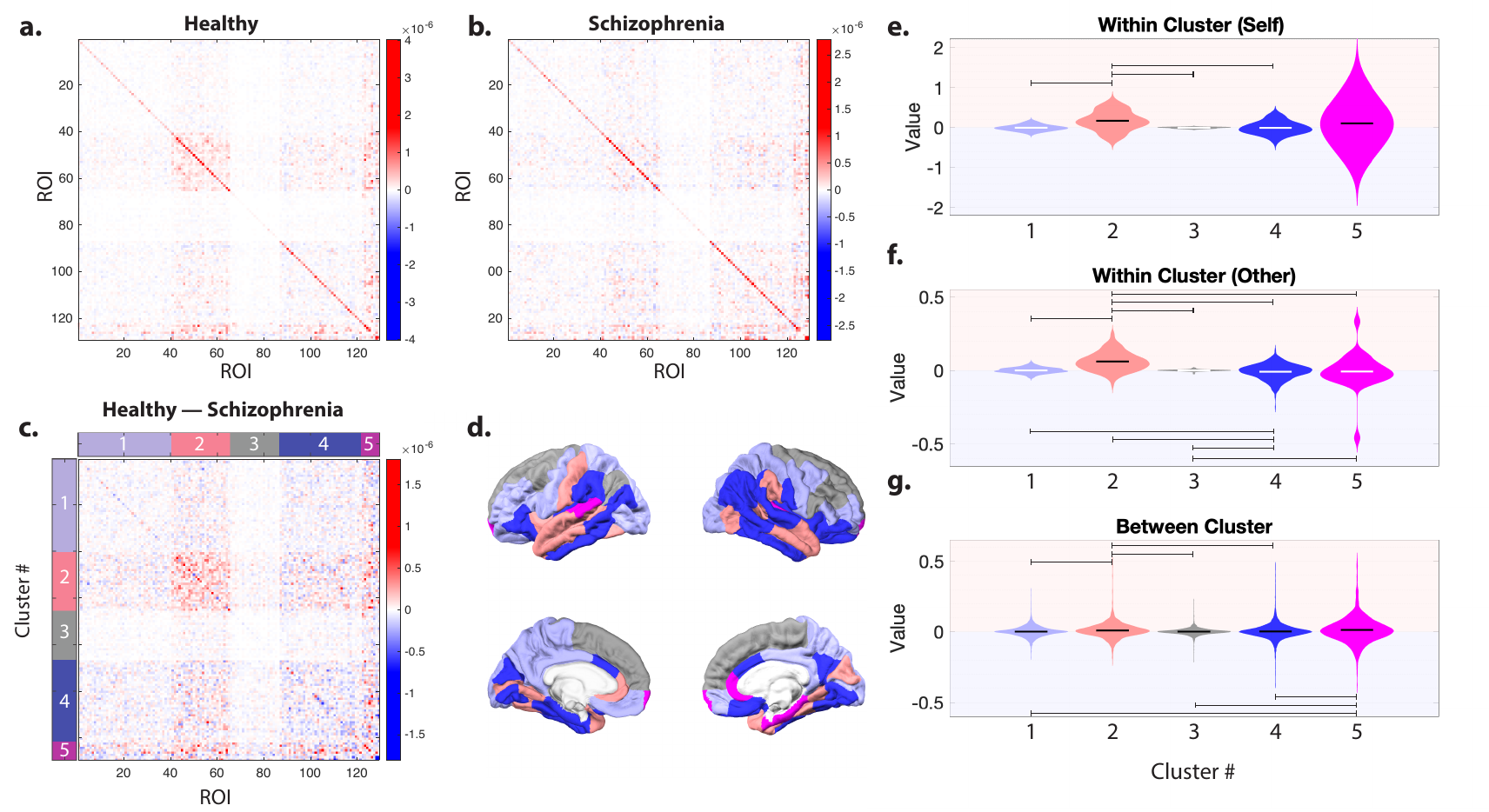}
\caption{\textbf{Pathological changes in the SC-FC coupling rules}. The group-average rule matrices $O$ for healthy controls \textbf{(a)} and individuals with SZ \textbf{(b)}. \textbf{(c)} Difference matrix (healthy minus SZ), with rows and columns ordered by WSBM-derived cluster assignments ($k=5$). \textbf{(d)} Brain surface overlay depicting the spatial distribution of the five WSBM clusters shown in panel \textbf{c}. Distributions of rule values extracted from the difference matrix in panel \textbf{c}: \textbf{(e)} within-cluster self-connections (diagonal elements), \textbf{(f)} within-cluster off-diagonal connections, and \textbf{(g)} between-cluster connections. Violin plots are color-coded by the WSBM community; black bars indicate the mean of positive-valued distributions, and white bars indicate the mean of negative-valued distributions. }
\label{Figure_6}
\end{figure*}

\subsection{Pathological changes in the structure-function coupling rules}

Changes in the coupling between SC and FC have been observed across various neurological and psychiatric conditions, including schizophrenia (SZ) \citep{zhao2023structure}, Major Depressive Disorder \citep{liao2024changes}, and Autism Spectrum Disorder \citep{qing2024structure}. In SZ, SC-FC coupling exhibits a complex and regionally specific pattern of alterations, with some brain networks demonstrating increased coupling while others exhibit decreased coupling \cite{sun2017modular}. These disruptions are thought to contribute to the aberrant neural dynamics underlying the cognitive and clinical symptoms characteristic of the disorder. 
Therefore, we hypothesized that our framework can identify the structural determinants of FC alterations by analyzing the estimated rule matrix $O$. 

To test this hypothesis, we utilized a publicly available structural and functional connectome dataset comprising SZ patients and healthy control participants -- for details on the dataset see the Materials and Methods section. We then assessed differences in the estimated rule matrices across SZ and control groups. Figure \ref{Figure_6} presents the average rule matrix $O$ for both the healthy control and SZ groups. Organizing SZ versus healthy control differences in the group-average rule matrices with a WSBM method highlights pronounced spatial heterogeneity in SC–FC coupling. Choosing $k=5$ blocks maximizes interpretability of between-group contrasts (Fig. ~\ref{Figure_6}c). The strongest alterations cluster within the temporal lobe. Cluster 2, which carries the largest positive rule weights in controls, is significantly attenuated in SZ (Wilcoxon rank-sum, $p<0.05$ versus nearly every other cluster). By contrast, Cluster 4 shows the opposite trend, with rule weights shifting toward more negative values in SZ. 

Finally, a compact bilateral set of ROIs, including the auditory cortices, frontal pole, right anterior cingulate cortex, and medial temporal regions, exhibits the greatest absolute deviations in rule weights across the brain, underscoring focal SC disruptions that may drive widespread functional alterations in SZ. Together, these findings demonstrate how our framework can uncover structured, network-level alterations in SC–FC coupling, providing a powerful lens to study the systems-level impact of brain disorders such as schizophrenia.

\section{Discussion}\label{sec12}

The intricate relationship between structural and functional connectivity in the human brain remains a central focus in network neuroscience and computational neuroimaging. While previous approaches have either relied on purely statistical correlations or complex biophysical simulations with limited interpretability, our generative linear framework offers a systematic approach to elucidate the topological and dynamical principles governing this relationship. 

By integrating DWI-derived SC and resting-state fMRI FC, our study contributes to the ongoing efforts to map the structural underpinnings of functional communication in the human brain. Specifically, our results provide mechanistic insights into structural determinants of FC through both direct and monosynaptic pathways and reveal organizational roles wherein specific regions act as structural \textit{linchpins} and \textit{fulcrums}, supporting global functional dynamics.

\subsection{Mechanistic insights from structure-function coupling rules}

\subsubsection*{The role of structural motifs in FC} 
The rule matrix $O$ in our framework provides a data-driven characterization of how direct and first-order indirect structural connections, along with mono-synaptic motifs, shape FC across the brain. The prominence of off-diagonal elements in the rule matrix suggests that more complex interactions, such as mono-synaptic motifs involving multiple regions, are essential to explain the observed FC beyond what can be accounted for by direct and first-order indirect SC alone.

These interactions establish the structural scaffolding for FC to emerge through long-distance synchronization among multiple regions, a mechanism theorized in communication models emphasizing path ensembles and polysynaptic signaling \citep{avena2017path,avenakoenigsberger2018communication,goni2014resting}. For instance, the dorsal and ventral attention networks, along with somatomotor regions, emerged as dominant contributors to functional dynamics supported by these structural motifs, aligning with attention networks' role in coordinating spatially distributed information processing \citep{corbetta2002control}.

\subsubsection*{Cortical and subcortical regulation of functional anti-correlations} 
We provided evidence that anti-correlations or reduced positive FC were mediated indirectly through structural motifs mainly involving several cortical regions within networks, such as somatomotor and frontoparietal networks, and subcortical regions, including bilateral nucleus accumbens (NAc), globus pallidus (GP), and amygdala. 

Subcortical regions such as NAc and GP contribute to large-scale functional connectivity through GABAergic polysynaptic loops involving the thalamus and prefrontal cortex. These regions act as anatomical and functional bottlenecks, integrating diverse inputs and modulating network activity via inhibitory signaling, which may suppress or prioritize information flow across cortical systems \citep{bell2016subcortical,zhao2024mediodorsal}.

Experimental and clinical studies, particularly those using deep brain stimulation in neuropsychiatric disorders, suggest that perturbations to the NAc and globus pallidus internus (GPi) can alter FC profiles across cortico-subcortical networks \cite{figee2013deep}. These changes often manifest as amplification or disruption of positive correlations (e.g., hyperconnectivity in cortico-striatal circuits) or as reduced anti-correlations between networks, such as the default-mode and task-positive systems. 

For example, experimental modulation of the ventral pallidum has been shown to toggle large-scale dynamics between task-positive and default-mode network states \citep{klaassen2021ventral}. While direct causal evidence for cortical regions' on global FC modulation remains limited, studies of neuropsychiatric disorders such as depression provide evidence for increased FC to cortical regions, including frontoparietal networks implicated in cognitive control, reflecting impaired top-down regulatory control \citep{sheline2010resting, zhou2010increased}. 

\subsubsection*{Emergence of anti-correlations through network dynamics}

Despite strong evidence for the role of inhibitory mechanisms, large-scale modeling studies have shown that anti-correlations in FC can emerge even in the absence of explicitly negative structural connections. A variety of modeling frameworks -- including neural mass models, mean-field approximations, and dynamic mean-field models -- have demonstrated that factors such as time delays, oscillatory phase differences, and global coupling strength can induce anti-correlated activity between regions that are structurally connected through predominantly excitatory pathways \citep{deco2009key, cabral2011role, hellyer2014control}. 

These phenomena reflect emergent network-level dynamics, whereby ensembles of regions become functionally segregated through collective behavior rather than direct inhibitory wiring. Thus, caution is warranted when interpreting high-magnitude rule weights as simple direct proxies for the density of structural connections, or negative weights as definitive evidence of inhibitory projections. Instead, the rules should be inferred as capturing the net effect of multi-scale mechanisms, including microcircuit properties, mesoscale architectural constraints, and large-scale dynamical interactions that jointly shape the observed FC between distant brain regions. 

\subsubsection*{Primacy of direct structural connections over global topological features}

Despite an incomplete understanding of all underlying mechanisms, our results provide strong evidence for the importance of direct SC and monosynaptic structural motifs in shaping the observed FC. Specifically, we found that replacing SC with SC-derived matrices, such as search information, which quantifies the ease with which information can traverse the shortest path between two nodes, significantly reduces the model’s predictive performance. 

This indicates that incorporating shortest-path-based metrics, which integrate global topological information, may obscure or dilute the local, monosynaptic features that are critical for accurate FC prediction. These findings suggest that, unlike unbiased path-based features of SC, our framework's predictive power stems from its ability to estimate the heterogeneous and localized effects of direct SC connections across the brain for each subject in shaping FC. 

Taken together, our findings underscore that preserving the fine-grained, subject-specific architecture of direct anatomical links yields a markedly more faithful and mechanistically interpretable account of whole-brain functional dynamics than substituting those details with abstract path-based statistics.

\subsection{Generative models of SC-FC couplings}

\subsubsection*{Historical evolution of structure-function modeling approaches}

The relationship between SC and FC has been investigated through diverse methodological techniques, each providing unique insights into the fundamental principles that govern brain network organization \citep{fotiadis2024structure}. Early statistical approaches mapped the topology of structural and functional networks, revealing how the architecture of the connectome constrains large-scale functional interactions \citep{messe2014relating,mivsic2016network}. While these studies, and the more recent machine learning approaches \citep{sarwar2021structure,zhang2022predicting} established the structural connectome as a critical scaffold for FC, they offered limited insight into the dynamic mechanisms driving SC-FC coupling \citep{li2022descriptive}. 

Computational network neuroscience modeling bridges this gap by formalizing mathematical relationships between structural architecture and functional dynamics, encoding principles of signal propagation, network communication, and emergent synchronization to elucidate how static anatomical pathways give rise to transient, context-dependent functional interactions. 

Stochastic approaches, such as maximum entropy models, infer interaction patterns from observed functional correlations, identifying probable structural network configurations and physiological insights \citep{watanabe2013pairwise,ashourvan2021pairwise}. In contrast, deterministic dynamical systems approaches simulate how structural connectivity constraints and time-evolving neural activity \cite{demirtacs2019hierarchical,tanner2024multi}, providing mechanistic accounts of how static architecture shapes time-varying interactions. 

Biophysical models are extended dynamical systems with biologically meaningful state variables governing physiological processes. These models have demonstrated that the interplay between SC and functional dynamics is influenced by factors such as coupling strength, time delays, and noise, collectively shaping the brain's spontaneous activity \citep{breakspear2017dynamic,honey2007network,deco2009key,herzog2024neural}. 

Communication models, in contrast, operationalize SC-FC coupling through intuitive, interpretable rules such as shortest path routing, diffusion, navigation, and path ensembles \citep{crofts2009weighted,goni2014resting,avena2018communication}. However, these models often overlook heterogeneity in governing principles across brain regions and individuals, limiting their ability to explain variability in functional dynamics. Recent studies demonstrated the utility of a flexible framework for modeling anatomically-constrained communication~\citep{betzel2022multi}. 

\subsubsection*{Bridging interpretability and accuracy}

Our work introduces a generative, individualized framework that achieves high accuracy in predicting FC, while retaining high interpretability. Critically, we demonstrate how structural motifs, such as triangular connections, shape the balance between integration and segregation in global functional dynamics. These findings provide mechanistic insights into how localized anatomical interactions scale to system-wide phenomena. 

Furthermore, despite the large number of model parameters, SC null testing revealed that the model’s ability to accurately predict an individual's FC is highly specific to their own SC. This finding strengthens our confidence that the results reflect genuine coupling rules rather than overfitting. Together, these findings suggest that a simple monosynaptic communication model, which effectively captures intersubject variability and heterogeneity in coupling rules, can robustly explain the emergence of global FC patterns.

Our findings also show accurate FC predictions even when using a different subject's SC or a randomized SC null model, provided the model parameters are re-estimated. This suggests that aligning an individual's structural and functional connectomes isn't crucial for high predictive performance, as long as the SC's topological features are preserved.

Our findings also underscore the individual variability in SC-FC coupling by demonstrating the superior predictive power of subject-specific rules over group-level rules. This observation reinforces the findings from connectome identifiability \citep{finn2015functional,amico2018quest} and other research demonstrating that personalized connectome architecture accounts for unique functional dynamics \citep{cooper2024personalized}, suggesting that group-level models may overlook critical idiosyncrasies in brain organization.

\subsection{Clinical and translational impact}

\subsubsection*{Framework for forecasting functional impact of structural alterations}

Understanding how the effect of structural disconnections propagates through functional networks has direct implications for clinical applications \citep{luppi2023computational}. Our framework provides a principled method for quantifying the functional impact of focal structural lesions, such as stroke, traumatic brain injury, and neurodegenerative disorders.

Previous studies have underscored the critical role of white matter integrity in maintaining cognitive function, showing that disruptions in the coupling between structural and functional connectivity serve as key indicators of cognitive impairment \citep{popp2024structural,wang2018alterations} and pathological conditions \citep{zhang2021relationship,ma2021selective}. 

Our findings in the schizophrenia dataset reveal a clear disruption in the normal coupling patterns between structural and functional connectivity, particularly within temporo-limbic and frontoparietal circuits. These alterations are consistent with prior evidence showing that SC in SZ is characterized by reduced white matter integrity, particularly in associative tracts, such as the Cingulum bundle, Uncinate fasciculus, and Superior longitudinal fasciculus, linking the frontal and temporal lobes \citep{kubicki2007review,dabiri2022neuroimaging}. A pivotal finding of our study is that alterations in SC in pathology are not the sole determinants of observed pathological FC. Instead, our results compellingly demonstrate that the SC-FC coupling rules themselves undergo changes in the SZ group.

This rule alteration can be conceptualized by considering that if SC between a pair of regions is diminished or absent in pathology, the algorithm will inherently assign minimal rule weight (especially with regularization) to predict their FC. Conversely, an increase in certain rule weights indicates a magnified contribution from the associated SC, potentially reflecting either an actual increase in SC or, critically, pathological compensatory mechanisms that lead to an observed increase in FC. Although a comprehensive elucidation of the neurobiological meaning of these \mbox{SZ-related} changes in FC is beyond the present scope, our work establishes a new frontier for exploring how SC-FC coupling rules are impacted by aging or various pathologies. 

\subsubsection*{Applications in neurosurgical planning}
In addition, our model uniquely extends current understanding by offering predictive insights into the causal links between specific SC disruptions and large-scale FC reorganization. One highly promising application lies in preoperative neurosurgical planning. Virtual lesion experiments have been used to estimate post-surgical functional outcomes in patients with epilepsy \citep{jirsa2023personalised}, stroke \citep{idesis2024generative}, and tumor resection patients \citep{aerts2020modeling}. Our generative model holds the potential to refine predictions by explicitly accounting for subject-specific SC-FC relationships, leading to enhanced prognostic accuracy. 

By systematically comparing SC-FC coupling across a range of neurological and psychiatric disorders, future research could pinpoint disorder-specific alterations in structure-function rules. This capability would be instrumental in accurately predicting changes in or recovery of FC, ultimately facilitating biomarker discovery and the development of individualized treatment strategies.

\subsection{Limitations and future directions}

\subsubsection*{Linearity assumption and omitted higher-order effects}

Despite its strengths, our generative model has several limitations. First, the assumption of linearity in SC-FC relationships, while computationally tractable, may oversimplify the nonlinear relationships, for example, between the fiber density and functional correlation, that underlie large-scale neural communication. Second, while our model incorporates indirect monosynaptic SC influences, it does not account for higher-order SC (e.g., \citep{goni2014resting}). Future work should compare our framework forecast power against models with higher-order factors to establish the importance of higher-order SC connections. 

\subsubsection*{Challenges in validating personalized SC-FC rules}

The observed lower accuracy in predicted FC using the group model likely reflects intersubject variability in the identified rules. This discrepancy may also arise from differences between group-level and subject-level estimation methods. However, it's important to consider that this variation could indicate subject-level overfitting despite our regularization efforts. This potential overfitting is a direct consequence of our framework's strength, which is providing accurate, individualized FC predictions through a data-driven generative approach. Therefore, to validate the subject-level rules, repeated measurements or causal studies, such as predicting post-surgical functional dynamics, are essential. It's important to acknowledge that the stability of SC-FC coupling rules over time or following surgical intervention is a confounding assumption for these approaches. 

\subsubsection*{Toward modeling the dynamic support of FC by SC}

In this work, we explored an alternative strategy for probing generalization to unseen subjects' FC using the BHA2 dataset, which provided around 15 minutes of resting-state time series. Our findings indicate that substantially longer time series are necessary to determine an FC estimation window size that promotes generalizability across unseen FC. Despite this, we observed that shorter windows, while generally showing poor generalizability, yielded highly accurate predicted FC in the training dataset. 

This suggests that while static SC-FC mapping may be achievable, long-duration correlation-based FC might not fully capture the time-varying support of FC by SC. Instead, our results imply that SC may precisely explain how dynamic brain activity is supported and emerges from SC. This would necessitate a time-varying $O$ matrix, illustrating how specific SC components actively contribute to ongoing FC, directly or indirectly. This opens a promising research direction for investigating dynamic rule matrices and their role in explaining time-varying brain activity.


\subsubsection*{Incorporating regional microstructural and molecular factors}

In this work, for the dataset without information on ROI self-connection fiber counts (i.e., BHA2), we assumed a uniform value for the structural self-coupling parameters across all brain regions. However, future work can improve the prediction of our model by informing these parameters by integrating regional information that affects functional activity. For example, local microstructural integrity metrics measured with advanced diffusion MRI metrics such as Neurite Orientation Dispersion and Density Imaging \citep{kraguljac2023neurite}, or molecular influences through genetic \citep{hawrylycz2012anatomically} and neurotransmitter profiling~\citep{hansen2022mapping}. 

\subsubsection*{Extending to directed and causal connectivity measures}

Finally, we examined correlation-based FC and demonstrated that SC can shed light on the origins and signs of certain spurious correlations. Future research should build upon this framework by incorporating effective connectivity measures, such as linear time-invariant models \citep{ashourvan2022external} and dynamic causal modeling \citep{friston2003dynamic}. These methods provide valuable insights into the directed and causal influences between brain regions and deepen our understanding of how the structural connectome drives functional dynamics.  

\subsection{Conclusion}

Our study introduces a novel generative framework that bridges statistical and mechanistic modeling approaches by formalizing structure-function coupling as an explicit mathematical relationship ($B = XOX^T$). This framework advances beyond existing methods by decomposing functional connectivity into interpretable structural rules that capture direct anatomical connections and monosynaptic motifs, while achieving high predictive accuracy through subject-specific parameterization.

The identification of brain regions as structural \textit{linchpins} and \textit{fulcrums} provides a new organizational taxonomy for understanding how local anatomical features scale to global functional dynamics. Our findings reveal that functional connectivity emerges primarily from direct structural connections rather than abstract global topological properties, challenging prevailing assumptions about polysynaptic communication.

Clinically, our framework offers a principled approach for predicting functional reorganization following structural disruptions. The demonstration that pathological changes involve reorganization of coupling \textit{rules} themselves -- not just structural alterations -- opens new avenues for precision medicine applications in neurology and psychiatry. By providing a mechanistic understanding of how specific structural disruptions give rise to particular functional deficits, this work lays the groundwork for the next generation of structure-informed interventions in brain health and disease.





\section{Materials and Methods}\label{sec11}
\subsection{Data and preprocessing}
We utilized publicly available datasets, providing subject-level structural and functional connectivity matrices derived from diffusion imaging and resting-state functional MRI.

The dataset that we analyzed in the main manuscript includes 50 healthy adults (23 female; mean age 29.5 ± 5.6 years; 47 right-handed) \citep{royer2022mridataset}. SC matrices were generated via anatomically constrained tractography and SIFT2-weighted streamlines \citep{smith2015sift2}, and FC matrices were computed by Pearson correlation of parcellated rs-fMRI time series. More details on the imaging protocol and data preprocessing can be found in  \citep{royer2022mridataset}.

We also leveraged a second dataset that consists of 27 healthy adults (35 ± 6.8 years) and 27 schizophrenia-spectrum patients (41 ± 9.6 years), all scanned at 3 T (Siemens Trio) and anonymized with ethics approval from the University of Lausanne \citep{Griffa2017,Daducci2012}. Precomputed SC and FC matrices were obtained from diffusion spectrum imaging and eyes-open resting-state fMRI, respectively, using the Connectome Mapper pipeline. The SC matrix elements represent the number of streamlines running between the regions, adjusted for average streamline length and regional sizes, and FC matrices were computed by Pearson correlation of parcellated rs-fMRI time series.

Finally, we used a third dataset -- Brain Hierarchical Atlas 2 (BHA2) \citep{jimenez2024open} -- that, in addition to SC and FC, provided subjects' resting-state time series, which we used to explore the effect of model parameters' sparsity on FC prediction accuracy (for more details, see the Subject-level fitting section). This dataset includes 136 healthy young adults (aged 20–30) from the MPI-LEMON cohort \citep{JimenezMarin2023,Babayan2019}. We used the provided SC and FC matrices constructed from preprocessed DWI and rs-fMRI data, using a 183-region parcellation from hierarchical clustering. Each entry in the SC matrices represents the number of stream counts between two regions, and each entry in the FC matrices represents the functional correlation between two regions.

\subsection{Group-level model solution}\label{groupLevelModel}

For the group-level fits, for each subject $i \in \{1, 2, \ldots, n\}$, we have the equation $\boldsymbol{B}_i=\boldsymbol{X}_i\boldsymbol{O}\boldsymbol{X}_i$, where $\boldsymbol{B}_i, \boldsymbol{X}_i, \boldsymbol{O} \in \mathbb{R}^{N \times N}$ represent the FC matrix, SC matrix, and rule matrix, respectively, with $N$ denoting the number of brain regions. Thus, the optimal rules matrix, at the group level, for the set of individuals is the matrix, $\boldsymbol{O}$, that minimizes

\begin{equation}
\label{}
\|\sum_{i} (\boldsymbol{X}_i\boldsymbol{O}\boldsymbol{X}_i - \boldsymbol{B}_i)\|_2.
\end{equation}

\noindent However, by definition, the rules matrix $\boldsymbol{O}$ must be symmetric. To ensure that the solution to this minimization problem is symmetric, we define $\boldsymbol{K} = \boldsymbol{X} \otimes \boldsymbol{X}$ and modify $\boldsymbol{K}$ and $\boldsymbol{O}$ using the following algorithm -- for a visual explanation, see Fig. 3 in the Supplementary Information:
\begin{enumerate}
    \item Let $L = \{(i, j) \in \mathbb{N} \times \mathbb{N} \mid i < j < n \}$, where $n$ denotes the number of rows (columns) of $\boldsymbol{X}$;
    \item Index $\boldsymbol{K}$ starting at 0;
    \item For each $(i, j) \in L$, add the $(in + j)$th column of $\boldsymbol{K}$ to the $(jn + i)$th column of $\boldsymbol{K}$;
    \item  For all $ (i, j) \in L$, delete the $(in + j)$th column of $\boldsymbol{K}$the corresponding element of $\text{vec}(\boldsymbol{O})$ simultaneously.
\end{enumerate}
\noindent Finally, denoting the transformed matrices as $\widetilde{\boldsymbol{K}}$ and  $\text{vec}(\widetilde{\boldsymbol{O}})$, we can write
\begin{equation}
\text{vec}(\widetilde{\boldsymbol{O}}) = 
\begin{pmatrix}
\widetilde{\boldsymbol{K}}_1\\ \widetilde{\boldsymbol{K}}_2 \\ \widetilde{\boldsymbol{K}}_3\\ \vdots \\ \widetilde{\boldsymbol{K}}_n
\end{pmatrix}^{\dagger}
\begin{pmatrix}
\text{vec}(\boldsymbol{B}_1) \\ \text{vec}(\boldsymbol{B}_2) \\ \text{vec}(\boldsymbol{B}_3)\\ \vdots \\ \text{vec}(\boldsymbol{B}_n)
\end{pmatrix},
\end{equation}
\noindent where $\dagger$ represents the Moore-Penrose pseudoinverse. No regularization is needed because the matrix formed by $\widetilde{\boldsymbol{K}}_1, ..., \widetilde{\boldsymbol{K}}_n$ was consistently found to be full rank. The matrix $\boldsymbol{O}$ can then be easily recovered, as $\widetilde{\boldsymbol{O}}$ simply contains the upper triangular elements of $\boldsymbol{O}$.




\subsection{Subject-level fitting}
For each subject-level fit, the equation $\boldsymbol{B}_i=\boldsymbol{X}_i\boldsymbol{O}_i\boldsymbol{X}_i$ is underdetermined. To account for this, we utilize LASSO regression \citep{Tibshirani1996}. The regularization parameter $\lambda \in \mathbb{R}^+$ controls the sparsity of the solution. The objective function, for each subject $i$, now becomes
\begin{equation}
\label{}
\frac{1}{2}\|(\boldsymbol{X}_i\boldsymbol{O}\boldsymbol{X}_i - \boldsymbol{B}_i)\|_2 + \lambda\|\boldsymbol{O}_i\|_1.
\end{equation}

After undergoing the same transformation used for the group-level fits, described in Section~\ref{groupLevelModel}, we can now rewrite the problem as $vec(\boldsymbol{B}_i) = \widetilde{\boldsymbol{K}}_i vec(\boldsymbol{O}_i) + \lambda|vec(\boldsymbol{O}_i)\|_1$.

\subsubsection*{Challenges in cross-validation for global models}
The regularization term, $\lambda$, dictates the sparsity of model parameters; a higher $\lambda$ value results in models with fewer active parameters. This promotes generalization and helps prevent overfitting, particularly in models with a large number of parameters. Ideally, $\lambda$ is selected via cross-validation on an independent test dataset. 

However, our model uniquely leverages the relationships between all variables to predict FC. This characteristic prevents the standard practice of splitting data into training and testing subsets, as it would disrupt the global variable relationships essential for our model's operation. 
Thus, traditional cross-validation would require datasets with repeated SC and FC measures, which were unavailable for this study.

\subsubsection*{Alternative generalizability assessment using BOLD time series}
Given the limitations of traditional cross-validation and recognizing the arbitrary nature and potential inconsistency of criteria like AIC and BIC, we pursued an alternative approach to evaluate model generalizability. We posited that repeated DWI scans, acquired within a short temporal window (hours), would not reflect actual changes in SC. Therefore, our focus was to assess the model's generalization to unseen FC. As the \cite{royer2022mridataset} dataset did not contain repeated FC measures, we utilized the BHA2 dataset. This dataset provided BOLD time series for ROIs, enabling the construction of distinct training and test FC matrices. Hence, it allowed us to investigate the impact of the FC estimation window length on model performance.

\subsubsection*{Regularization parameter effects on model performance}

We evaluated the effect of the $\lambda$ parameter on model performance using three sample subjects, presenting results for the training and test FC datasets in SI Fig. \ref{SI_Figure_7_SI}a and \ref{SI_Figure_7_SI}b, respectively. Consistent with expectations, higher $\lambda$ values led to sparser rule matrices and diminished model performance.

In the training dataset (SI Fig. \ref{SI_Figure_7_SI}a), models generally performed best with larger FC estimation windows, though short windows yielded high accuracy for two subjects. Performance consistently dropped when rule matrix density fell below 60-80\%. The test dataset (SI Fig. \ref{SI_Figure_7_SI}b) showed similar patterns, but with a more pronounced performance reduction for smaller window sizes.

\subsubsection*{Binary search strategy for regularization parameter selection}
For regularization to clearly enhance generalization, peak performance should occur at lower densities (i.e., where regularization has pruned parameters). While some peaks appeared at densities below full for window sizes under 274 seconds, the overall model performance in these cases was already poor. Thus, despite observing limited signs of generalizability, the evidence was insufficient to guide the selection of the regularization parameter in this study.

Consequently, the $\lambda$ was determined through a binary search, targeting a density of $0.8 \pm 0.01$ non-zero elements within the $O$ matrix. This specific density was chosen as it yielded acceptable performance on the training data and offered a degree of generalizability to unseen FC. This binary search was initialized with a high and low value set to $\lambda_{max} = \frac{1}{N}|\langle \boldsymbol{x}^{k}_i, vec(\boldsymbol{B_i})\rangle$ and  $\lambda_{min} = \frac{\lambda_{max}}{100000}$. This maximum lambda value represents the value at which the fitted $\boldsymbol{O}$ matrix becomes all zeroes \citep{friedman2010regularization}. A 0.01 tolerance for the density was used so that binary search could converge to a $\lambda$ value in a reasonable number of steps. 

\subsection{Model performance evaluation}

To evaluate the model’s ability to predict empirical FC, we performed a linear regression between the predicted and observed FC values. Specifically, we regressed the upper triangular elements of the predicted FC matrix against those of the actual FC matrix for each subject. An intercept term was included to account for potential offsets in predicted FC values. Model performance was quantified using the coefficient of determination ($R^2$), reflecting the proportion of variance in observed FC explained by the model predictions.

\subsection{Statistical analysis and null models}

For each subject's FC matrix, 100 null matrices were generated, preserving the original degree and strength distributions. Randomization followed the method introduced by Rubinov et al. in 2011 \citep{Rubinov2011}, using the Brain Connectivity Toolbox null\_model\_und\_signed function, with the weight frequency parameter set to 1 and the bin swaps parameter set to 10. Next, for each subject $i$, 100 null rule matrices were computed, one for each null FC matrix, using lasso regression with the same penalty term, $\lambda$, that was selected in training the given subject. Finally, at the group level, 100 null rule sets were computed as follows:
\begin{equation} 
\label{eq:model}
vec(\boldsymbol{O}_i) = 
\begin{pmatrix}
\widetilde{\boldsymbol{K}}_1\\ \widetilde{\boldsymbol{K}}_2 \\ \widetilde{\boldsymbol{K}}_3\\ ... \\ \widetilde{\boldsymbol{K}}_n\
\end{pmatrix}^\dagger
\begin{pmatrix}
vec(\boldsymbol{B}_{i1}) \\ vec(\boldsymbol{B}_{i2}) \\ vec(\boldsymbol{B}_{i3})\\ ... \\ vec(\boldsymbol{B}_{in})\
\end{pmatrix},
\end{equation}
 where each \(\widetilde{\boldsymbol{K}}_i\) is computed in the same manner as in equation 2, but where each \(\boldsymbol{B}_{ij}\) is instead the $j$th null FC matrix for subject $i$.

These null rules were then utilized to determine statistically significant rules at both the subject and group levels. A Wilcoxon signed rank test was conducted to compare the \((i, j)\)th element of the rules matrix with the corresponding elements in each of the 100 null rules matrices. FDR correction was then applied to the resultant $p$-values to correct for multiple comparisons. Finally, rules whose associated $p$-values were less than 0.05 were kept. The remaining rules were set to 0.

\subsection{Virtual resections}
For each region of interest, all structural connections not involving the region or system of interest were removed. For a given region or system of interest, constructed of some subset of the parcellated regions, all elements of $\boldsymbol{X}$ except for those of the form $\boldsymbol{X}_{ij}$ and $\boldsymbol{X}_{ji}$ were set to $0$, where $i$ belongs to the set of indices belonging to the region or system of interest and $j$ belongs to the set of all indices. This procedure was performed on the matrix $\boldsymbol{X}$ for each subject for each region of interest. We then calculated the element-wise mean of each subject's predicted FC under the subject's modified $\boldsymbol{X}$ matrix. Two sets of these predictions were made both under the subject and group-level rules.

\subsection{Search information measure}

Search information is a network-theoretic metric that quantifies how difficult it is to navigate between two nodes in a graph when only local information is available at each step. Originally introduced in Goñi \emph{et al.}~\citep{Goni2014}, it has been widely used in brain network analysis to evaluate communication efficiency along shortest paths between regions of interest.

Given a weighted, undirected network (e.g., a structural connectivity matrix), the search information $S_{ij}$ between nodes $i$ and $j$ is defined as the negative logarithm of the probability of successfully following a shortest path from $i$ to $j$ using only local knowledge of each node’s neighbors, i.e.,
\begin{equation}
S_{ij} = -\log_2 P_{ij},
\end{equation}
where $P_{ij}$ is the product of transition probabilities along the shortest path $\pi(i \rightarrow j)$. At each node $k$ along this path, the probability of selecting the correct next step $k \rightarrow k'$ is computed by normalizing the edge weights among $k$'s neighbors:
\begin{equation}
P_{ij} = \prod_{(k \rightarrow k') \in \pi(i \rightarrow j)} \frac{w_{kk'}}{\sum_{m \in \mathcal{N}(k)} w_{km}}.
\end{equation}

Here, $w_{kk'}$ is the weight of the edge between nodes $k$ and $k'$, and $\mathcal{N}(k)$ denotes the set of neighbors of node $k$. This formulation assumes that at each step, the traveler has no global map and chooses the correct neighbor based only on edge weights.

Search information thus reflects the \emph{navigational cost} of traversing the shortest path under local constraints. A low value indicates that the path is easily discoverable with local knowledge, while high values suggest that accurate navigation requires more information \citep{Goni2014}.

\subsection{Weighted stochastic block model}

We used a weighted stochastic block model (WSBM) to identify communities in our weighted, sparse rule matrix. Let the network be represented by its 
adjacency matrix, \(\boldsymbol{O} \in \mathbb{R}^{N\times N}\), where \(\boldsymbol{O}_{ij}\) denotes the weight of the rule from node \(j\) to node \(i\). We assume that each node \(i\) belongs to one of \(K\) distinct communities, denoted by the label \(z_i \in \{1,\dots,K\}\). 

In the traditional (binary) stochastic block model, an edge between nodes \(i\) and~\(j\) exists with probability \(\theta_{z_i z_j}\), where \(\theta_{rs}\) is the probability of connection between communities \(r\) and \(s\). Formally, for an unweighted (binary) network, the likelihood of observing \(\boldsymbol{O}\) under this model is given by

\begin{equation}
P\bigl(\boldsymbol{O} \,\big\vert\, \{\theta_{rs}\}, \{z_i\}\bigr)
\;=\;\prod_{i,j>i}\,\theta_{z_i z_j}^{\,\boldsymbol{O}_{ij}}\;\bigl(1-\theta_{z_i z_j}\bigr)^{\,1 - \boldsymbol{O}_{ij}},
\tag{1}
\end{equation}
where \(\boldsymbol{O}_{ij} \in \{0,1\}\). By maximizing this likelihood with respect to the community labels \(\{z_i\}\) and probabilities \(\{\theta_{rs}\}\), one can uncover an arrangement of nodes into communities that capture their underlying connection patterns. 

Many real-world networks, including our rule matrix, are weighted and potentially sparse. To accommodate real-valued edge weights, one can adopt an exponential family version of the block model. In general, the likelihood function can be expressed as
\begin{equation}
P\bigl(\boldsymbol{O} \,\big\vert\, \{\theta_{rs}\}, \{z_i\}\bigr)
\;\propto\;
\exp\!\Biggl(\,
\sum_{i,j}\;T\bigl(\boldsymbol{O}_{ij}\bigr)\;\cdot\;\eta\bigl(\theta_{z_i z_j}\bigr)
\Biggr),
\tag{2}
\end{equation}
where \(T(\cdot)\) are sufficient statistics for the chosen distribution (e.g., a Gaussian, log-normal, etc.), and \(\eta(\cdot)\) are their corresponding natural parameters. The parameters \(\theta_{z_i z_j}\) thus characterize how any two communities, \(z_i\) and \(z_j\), tend to connect in terms of both the existence and the magnitudes of edges.

In the WSBM, as in the classical SBM, each node’s community assignment \(\{z_i\}\) and the parameters \(\theta_{z_i z_j}\) fully specify the model. The key difference is that \(\theta_{z_i z_j}\) now describes the distribution of edge weights rather than just their presence or absence. Following the approach in \citep{aicher2013adapting,betzel2018diversity}, we assume these weights are drawn from a normal distribution with sufficient statistics \(T = (x, x^2, 1)\) and natural parameters \(\eta = \bigl(\tfrac{\eta}{\sigma^2}, -\tfrac{1}{2\sigma^2}, -\tfrac{\mu^2}{2\sigma^2}\bigr)\). Thus, an edge between communities \(z_i\) and \(z_j\) is characterized by a mean \(\mu_{z_i z_j}\) and variance \(\sigma^2_{z_i z_j}\). The likelihood can then be written as

\[
P\bigl(\boldsymbol{A} \,\big\vert\, \{z_i\}, \{\mu_{rs}\}, \{\sigma^2_{rs}\}\bigr)
\;=\;
\prod_{i,j}\,
\exp\Bigl(
\,\boldsymbol{A}_{ij}\,\tfrac{\mu_{z_i z_j}}{\sigma^2_{z_i z_j}}
-\tfrac{\boldsymbol{A}_{ij}^2}{2\,\sigma^2_{z_i z_j}}
-\tfrac{\mu_{z_i z_j}^2}{2\,\sigma^2_{z_i z_j}}
\Bigr).
\]

We maximize the likelihood of this sparse WSBM using a Variational Bayes technique described in \citep{aicher2013adapting} and implemented in MATLAB using code (WSBM V1.2) made available on the author’s personal website (\url{https://aaronclauset.github.io/wsbm/}). This inference procedure jointly estimates the parameters \(\{\theta_{rs}\}\) for each pair of communities and the assignment \(\{z_i\}\) of each node, thereby offering a flexible framework for detecting multiple forms of mesoscale organization in weighted networks.

Due to the non-deterministic nature of the algorithm, we performed 10 independent runs with randomized initializations and selected partitions, maximizing the log evidence. To enhance interpretability, we selected a partition ($k=4$) for the manuscript figure that visually emphasized intuitive mesoscale rules (e.g., clear core-periphery distinctions). While our analysis focused on four clusters, future work should systematically characterize subject-specific rule matrices across different clustering resolutions. Such an exploration could deepen understanding of the mesoscale diversity of structural rules.

\section{Data availability statement}

All data used in this study are publicly available and can be accessed following the instructions provided in the Materials and Methods section.

\section{Code availability statement}

The scripts used to generate the results can be found on GitHub at \url{https://github.com/samkelemen/linear_generative_model}.

\section{Acknowledgments}
This work was supported by startup funding from the Department of Psychology at the University of Kansas provided to A.A.

\bibstyle{sn-nature} 
\bibliography{sn-bibliography_SP}
\newpage

\appendix

\section{Supplementary Figures}

\newpage

\begin{figure*}
\centering
\setcounter{figure}{0}

\includegraphics[width=1\linewidth ]{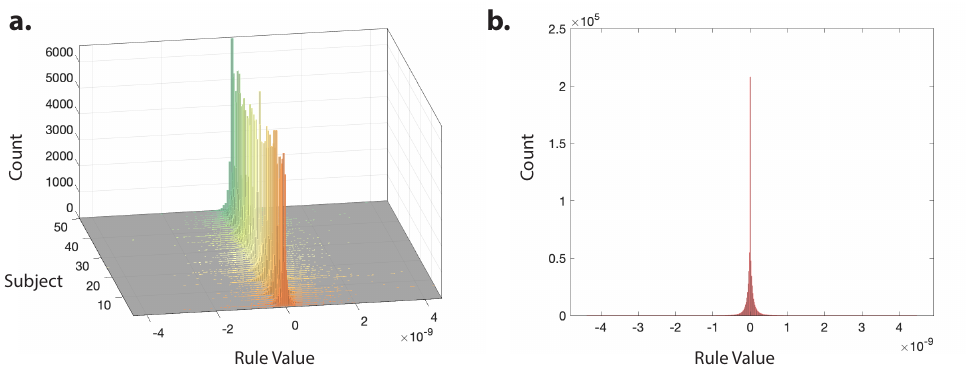}
\caption{\textbf{Distribution of rule values}\textbf{(a)} Distributions of subject-level rule matrix element values, color-coded for each subject. \textbf{(b)} Distribution of subject-level rule matrix element values aggregated across all subjects.}
\label{SI_Figure_1}
\end{figure*}

\begin{figure*}
\centering
\includegraphics[width=1\linewidth ]{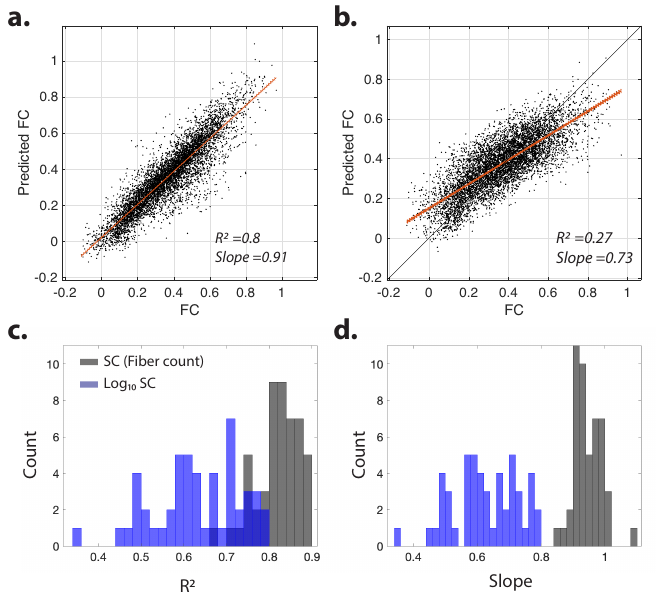}
\caption{\textbf{Effect of Log-transforming SC on the prediction of FC} \textbf{(a)} Scatter plot comparing the sample subject's actual FC values with the predicted FC values using the subject-specific rule matrix derived from the SC matrix of fiber counts between brain regions. The red line represents the linear fit, while the dashed line indicates the 95\% confidence interval. \textbf{(b)} Same as panel \textbf{(a)}, except we used $\log_{10}(\text{SC})$. To avoid issues with zero values, we added 1 to all elements of the SC matrix before applying the logarithm. \textbf{(c)} Distribution of $R^2$ values across all subjects ($n=50$) for the linear regression between actual and predicted FC values using subject-specific rule matrices derived from fiber count SC (black) and $\log_{10}(\text{SC})$ (blue). Predictions using $\log_{10}(\text{SC})$ result in a significant reduction in the mean goodness-of-fit ($R^2$) of the linear regression ($t$-test, $p=4.8 \times 10^{-18}$). \textbf{(d)} Distribution of slope values across all subjects ($n=50$) for the linear regression between actual and predicted FC values using subject-specific rule matrices derived from fiber count SC (black) and $\log_{10}(\text{SC})$ (blue). Predictions using $\log_{10}(\text{SC})$ lead to a significant reduction in the mean slope of the linear regression ($t$-test, $p=2 \times 10^{-38}$).}
\label{SI_Figure_2_Log}
\end{figure*}

\begin{figure*}
\centering
\includegraphics[width=1\linewidth ]{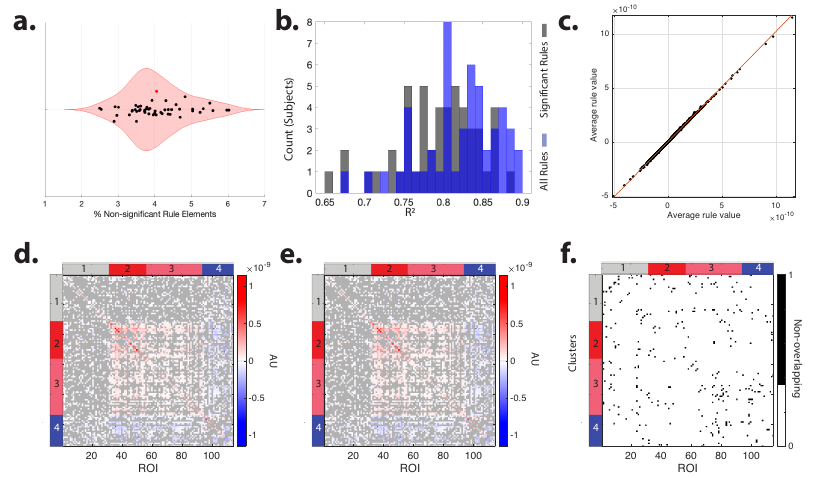}
\caption{\textbf{FC prediction using subject's unique rules} 
\textbf{(a)} The percentage of non-significant rules (nonparametric permutation testing using randomized FC nulls, $p<0.05$, FDR-corrected for multiple comparisons -- see  Materials and Methods section for details) relative to the total number of rules for each subject. The red dot shows the average percentage across all subjects. \textbf{(b)} Distributions of the $R^2$ values of the linear fit between the actual and predicted FC values using all subject-level rules (black) and using only the significant rule matrix elements (blue). Removing non-significant rules that are not significant for each subject results in a significant ($t-$test, $p=1.01 \times 10^-14$) reduction in the mean goodness-of-fit of a linear fit ($R^2$), comparing the actual versus predicted FC values. \textbf{(c)} The group-average subject-level rule values against the rule values after removing each subject's non-significant rules. Red line shows the linear fit (slope $=1.02$, $R^2=0.99$)  \textbf{(d)} Subject-level rule matrix $O$ averaged across all subjects, with brain regions sorted by their cluster assignments identified using the WSBM method ($k=4$). The rule elements with means that show no significant ($t-$test, $p<0.05$, FDR corrected for multiple comparisons across all rules) difference from zero across all subjects are color-coded in gray. \textbf{(e)} Same matrix in panel \textbf{d}, except the non-significant subject-level rules were removed before the calculation of group means. \textbf{(f)} The overlap between the matrices in panels \textbf{d} and \textbf{d} marked by white and black.}
\label{SI_Figure_2}
\end{figure*}

\begin{figure*}
\centering
\includegraphics[width=1\linewidth ]{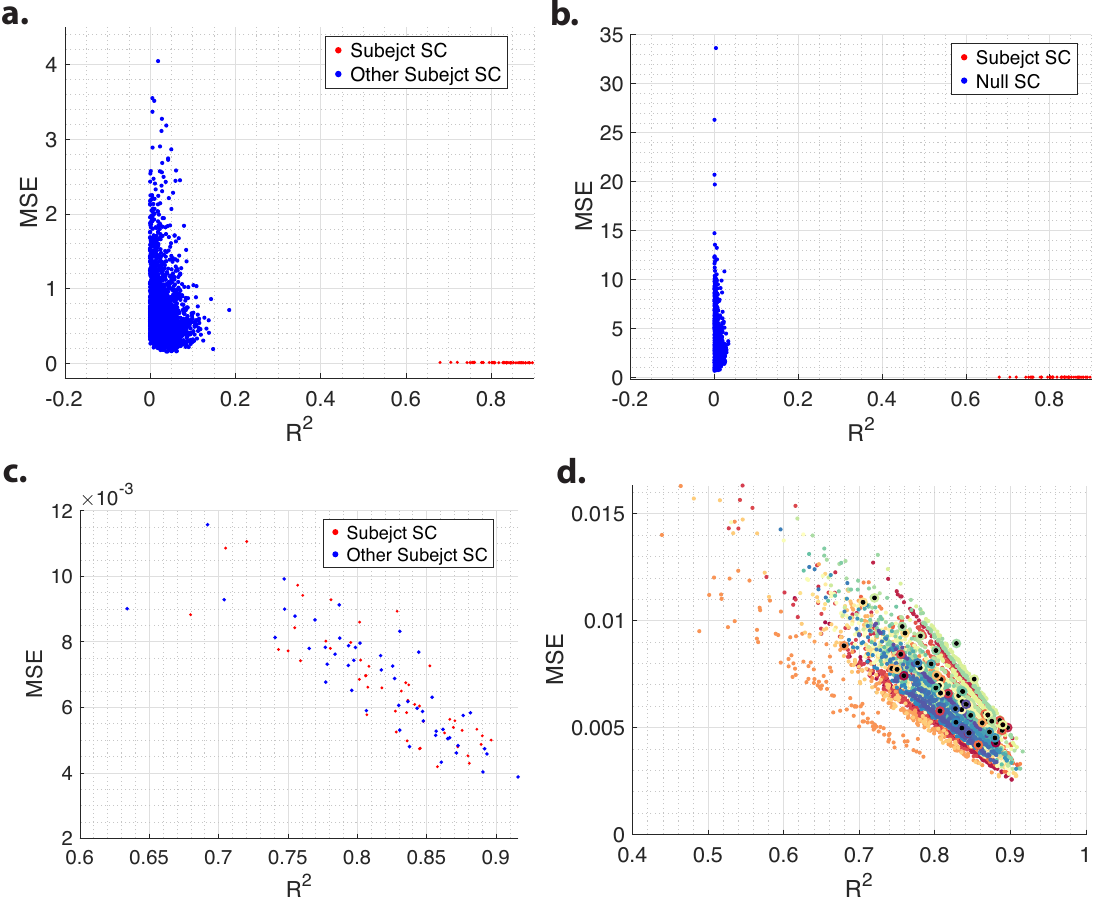}
\caption{\textbf{FC prediction using other subjects and null SC } 
\textbf{(a)} Goodness-of-fit for the linear regression between predicted and actual FC, evaluated using mean squared error (MSE) and $R^2$ values. Predictions based on each subject's SC are shown in red, while those using another subject's SC and the original subject's estimated rule matrix $O$ are shown in blue. \textbf{(b)} Same as panel \textbf{a}, except the blue dots represent predictions generated from SC randomized null models ($n=100$ nulls per subject) \textbf{(c)} Same setup as panel \textbf{a}, but here the rule matrix was re-estimated using a randomly selected subject’s SC to predict the original subject’s FC (blue dots). \textbf{(d)} Similar to panel \textbf{b}, except rule matrices were re-estimated for each SC randomized null model as in panel \textbf{c}. Subjects are color-coded, and actual predictions based on each subject’s own SC and FC are indicated with larger markers featuring a black center.}
\label{SI_Figure_3_new}
\end{figure*}

\begin{figure*}
\centering
\includegraphics[width=1\linewidth ]{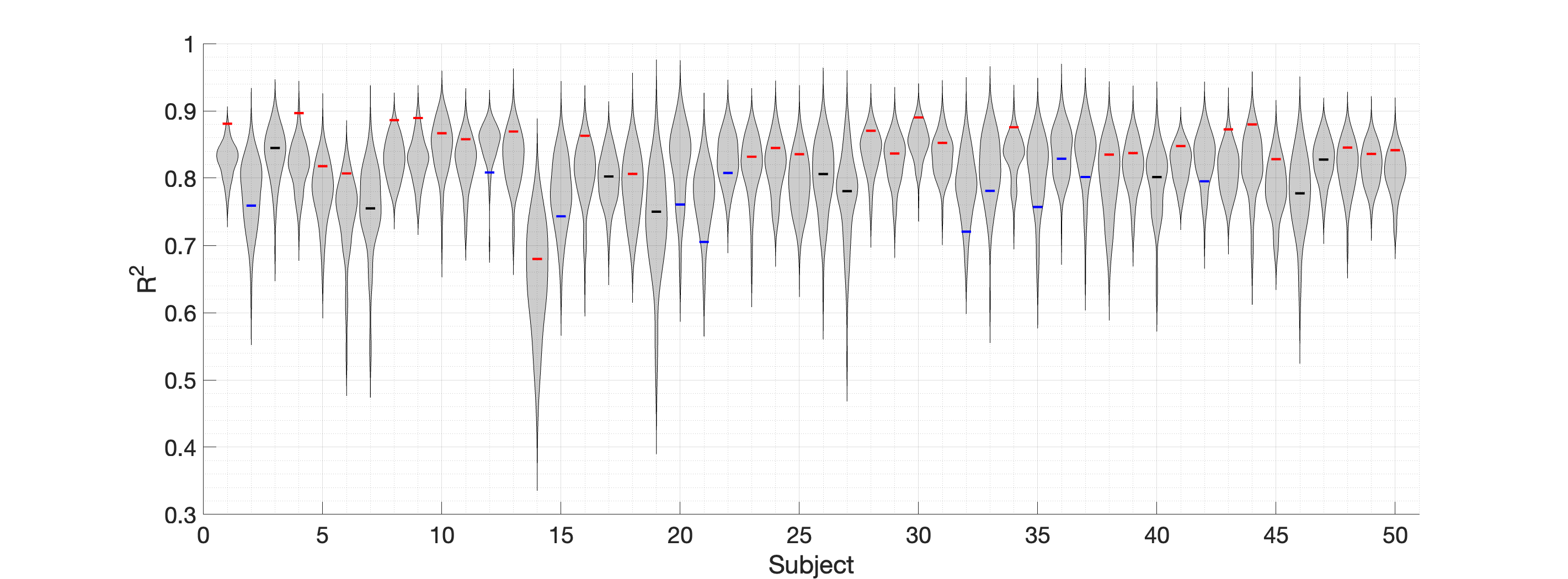}
\caption{\textbf{FC prediction using null SC } 
\textbf{(a)} Goodness-of-fit for the linear regression of actual versus predicted FC, assessed using $R^2$ values. $R^2$ values for the Predictions based on each subject's own SC are shown using the bar, and the violin plots show the $R^2$ values for predictions based on SC randomized null models, with rule matrices estimated from these SC nulls. The significantly ($t-$test,$p<0.05$, FDR corrected for multiple comparisons across subjects) higher or lower $R^2$ values than predicted based on SC randomized null models are color-colored in red and blue, respectively.}
\label{SI_Figure_3}
\end{figure*}


\begin{figure*}
\centering
\includegraphics[width=0.8\linewidth ]{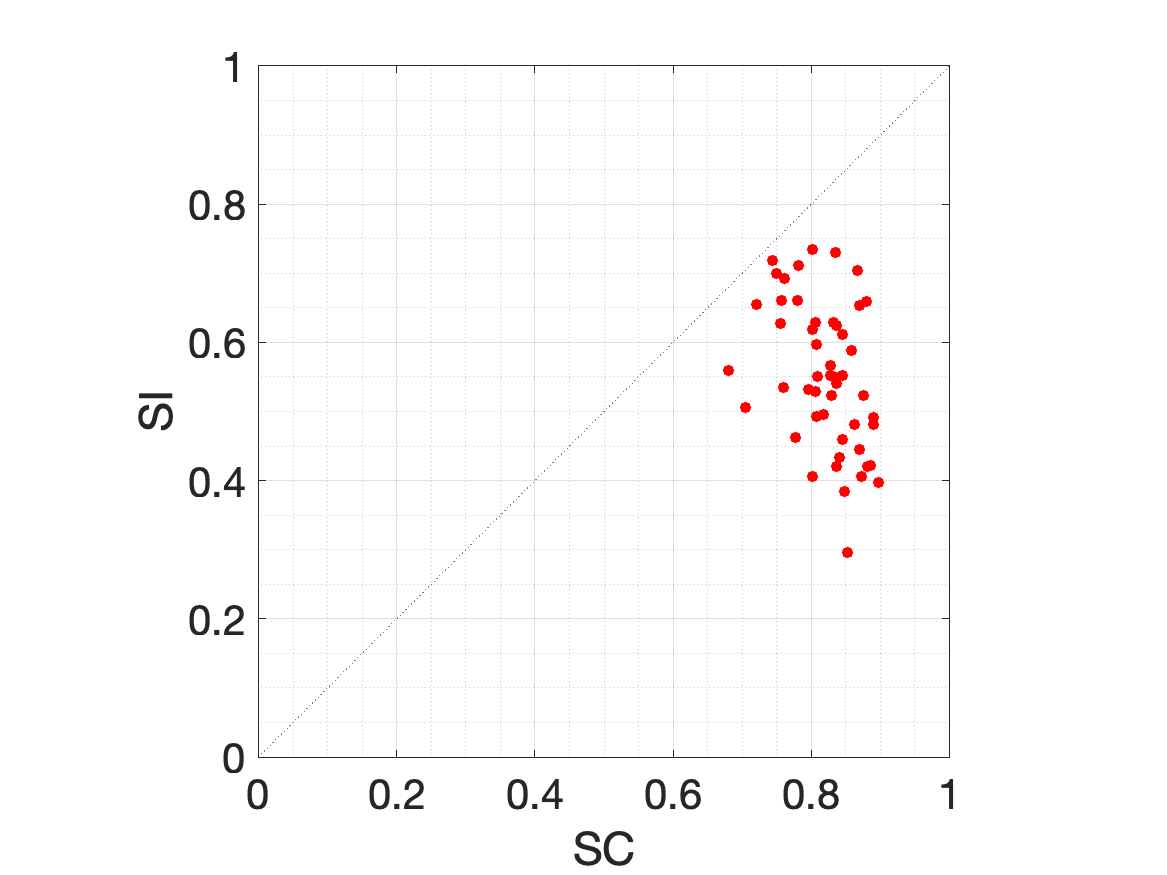}
\caption{\textbf{Search Information versus structural connectome prediction accuracy.} The plot represents the goodness-of-fit for the linear regression of actual versus predicted FC, assessed using $R^2$ values, for the Predictions based on each subject's own SC compared to those using the search information matrix derived from the subject's SC. The $R^2$ values are significantly higher when subjects' SC was used compared to the search information matrix to predict FC matrices ($t-$test, $p=8.65 \times 10^-19$).}
\label{SI_Figure_4_SI}
\end{figure*}

\begin{figure*}
\centering
\includegraphics[width=1\linewidth ]{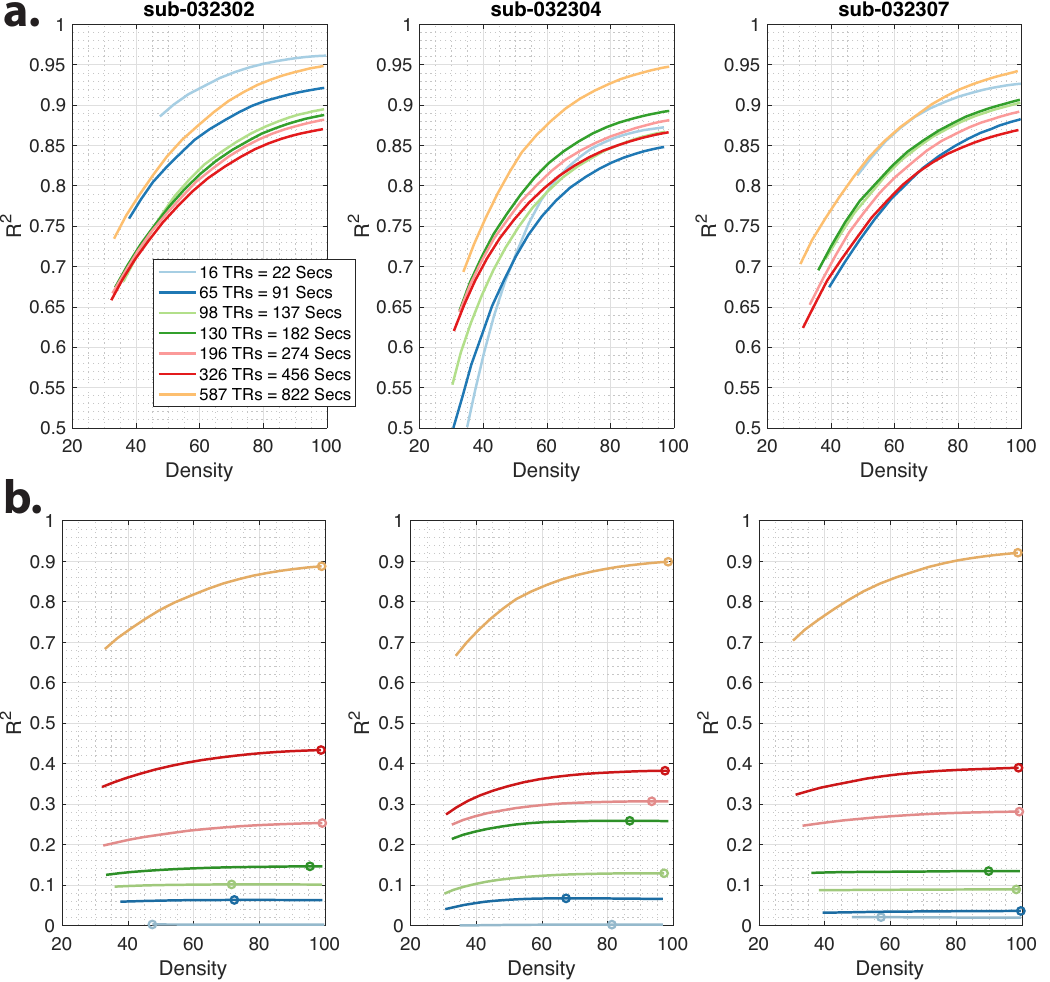}
\caption{\textbf{The effect of rule matrix density on prediction accuracy.} \textbf{(a)} Prediction accuracy ($R^2$) of the linear‐regression model versus rule‐matrix sparsity (percentage of nonzero entries), shown for color-coded window sizes corresponding to approximately 2 \%, 10 \%, 15 \%, 20 \%, 30 \%, 50 \% and 90 \% of the time series length. \textbf{(b)} Same as panel \textbf{a}, but prediction accuracy is evaluated against FC matrices computed from unseen (test) time series. Maximum values are marked by `o'. Note that the 90 \% window size overlaps with the training time series and is the only condition not evaluated on fully unseen data.}
\label{SI_Figure_7_SI}
\end{figure*}



\begin{figure*}
\centering
\includegraphics[width=1\linewidth ]{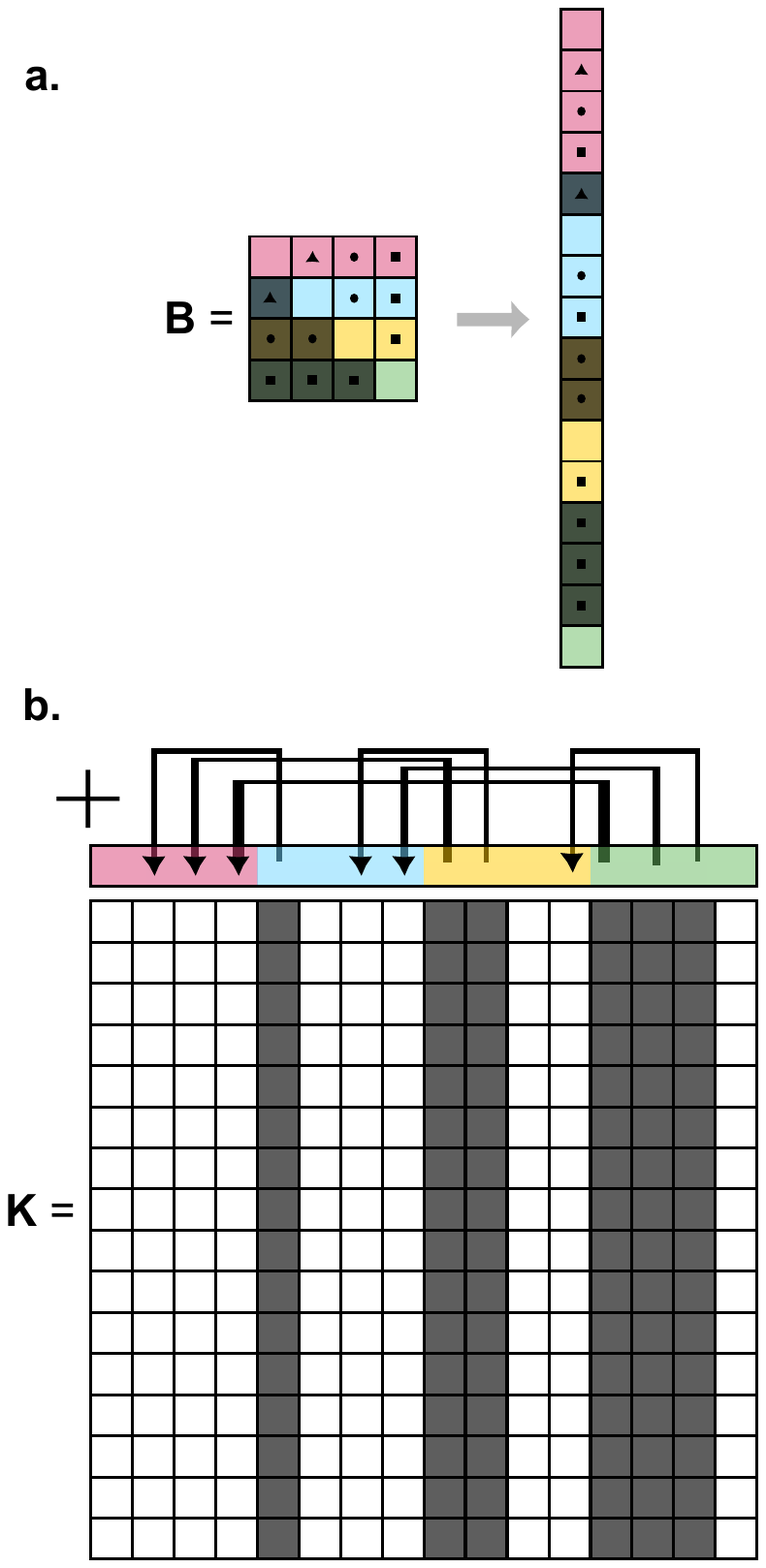}
\caption{\textbf{schematic of algorithm to preserve symmetry} The above figure is a visual representation of the algorithm in section 4.2 used to modify $\boldsymbol{K}$ and $\boldsymbol{B}$. \textbf{(a)} $\boldsymbol{B}$ undergoes the transformation $vec(\boldsymbol{O})$, then the grayed values are removed. \textbf{(b)} Each grayed column of $\boldsymbol{K}$ is added to the column to which its arrow points. The grayed columns are then deleted from $\boldsymbol{K}$. }
\label{SI_Figure_5}
\end{figure*}

\end{document}